\documentclass[letter,useAMS,usenatbib]{mn2e}

\usepackage[english]{babel} \usepackage{subfigure}
\usepackage{graphicx}

\usepackage[fleqn]{amsmath}
\usepackage{color}

\usepackage[varg]{txfonts}

\newcommand{\noi}{\noindent}

\title[Strong lensing on adaptive grids]{Bayesian strong gravitational-lens modelling on adaptive
grids:\\ objective detection of mass substructure in galaxies}

\author[S. Vegetti \& L. V. E.  Koopmans.]{ Simona Vegetti\thanks{E-mail:
    vegetti@astro.rug.nl} \& L. V. E.  Koopmans\\ Kapteyn
    Astronomical Institute, University of Groningen, PO Box 800,
    9700\,AV Groningen, the Netherlands}

\begin{document}
  
  \date{Accepted for publication on MNRAS}
  
  \pagerange{\pageref{firstpage}--\pageref{lastpage}} \pubyear{2008}
  
  \maketitle
  
  \label{firstpage}
  
  \begin{abstract}
    
    We introduce a new adaptive and fully Bayesian grid-based method
    to model strong gravitational lenses with extended images. The
    primary goal of this method is to quantify the level of luminous
    and dark-mass substructure in massive galaxies, through their
    effect on highly-magnified arcs and Einstein rings. The method is
    adaptive on the source plane, where a Delaunay tessellation is
    defined according to the lens mapping of a regular grid onto the
    source plane. The Bayesian penalty function allows us to recover
    the best non-linear potential-model parameters and/or a grid-based
    potential correction and to objectively quantify the level of
    regularization for both the source and the potential. In addition,
    we implement a Nested-Sampling technique to quantify the
    errors on all non-linear mass model parameters -- marginalized
    over all source and regularization parameters -- and allow an
    objective ranking of different potential models in terms of the
    marginalized evidence. In particular, we are interested in
    comparing very smooth lens mass models with ones that
    contain mass-substructures. The algorithm has been tested on a range
    of simulated data sets, created from a model of a realistic
    lens system. One of the lens systems is characterized by a smooth
    potential with a power-law density profile, twelve 
    include a Navarro, Frenk and White (NFW) dark-matter substructure of different masses and at
    different positions and one contains two NFW dark substructures with the 
    same mass but with different positions. 
    Reconstruction of the source and of the lens
    potential for all of these systems shows the method is able, in a
    realistic scenario, to identify perturbations with masses $\ga 10^7\rm
    M_\odot$ when located \emph{on} the Einstein ring. For
    positions both inside and outside of the ring, masses of at least
    $10^9\rm M_\odot$ are required (i.e. roughly the Einstein ring of
    the perturber needs to overlap with that of the main lens). Our
    method provides a fully novel and objective test of mass
    substructure in massive galaxies.

\end{abstract}
  

  \section{Introduction}
  
  At the present time, the most popular cosmological model for
  structure formation is the $\Lambda \text{CDM}$ paradigm. While this
  model has been very successful in describing the Universe on large
  scales and in reproducing numerous observational results
  \citep[e.g.,][]{Reiss98, Efstathiou02, Burles01,
  Philips01, Jaffe01, Percival01, deBernardis02, Hamilton02, Croft02,
  Tonry03, Spergel03, Komatsu08}, important discrepancies still
  persist on small scales. In particular, some of these involve the
  dark matter distribution within galactic haloes
  \citep[e.g.,][]{Moore94, Burkert95, McGaugh98,
 Binney01, Blok01, deBlok02, McGaugh03, Simon03,Rhee04,Kuzio06} 
 and the number of galaxy satellites, i.e the
  \emph{Missing Satellite Problem}.
  
  \noi According to the standard scenario, structures form in a
  hierarchical fashion via merging and accretion of smaller objects
  \citep{Toomre77, Frenk88, White91, Barnes92, Cole00}. As shown by
  the latest numerical simulations, in which high mass and force
  resolution is achieved, the progenitor population is only weakly
  affected by virialization processes and a large number of sub-haloes
  is able to survive after merging. The number of substructures
  within the Local Group, however, is predicted to be 1-2 orders of
  magnitude higher than what is effectively observed
  \citep[e.g.,][]{Kauffmann93, Moore99, Klypin99,
  Moore01,Diemand07b,Diemand07a}.
  
  \noi Two different classes of solutions have been suggested to
  alleviate this problem, cosmological and astrophysical. Cosmological
  solutions address the basis of the $\Lambda \text{CDM}$ paradigm
  itself and mostly concentrate on the properties of the dark matter,
  allowing for example, for a warm \citep{Colin00}, decaying
  \citep{Cen01}, self-interacting \citep{Spergel00}, repulsive
  \citep{Goodman00}, or annihilating nature
  \citep{Riotto00}. Alternatively the $\Lambda \text{CDM}$ picture can
  be modified by the introduction of a break of the power-spectrum at
  the small scales \citep[e.g.,][]{Kamionkowski00, Zentner03}.
  
  \noi From an astrophysical point of view, the number of visible
  satellites can be reduced by suppressing the gas collapse/cooling
  \citep[e.g.,][]{Bullock00, Kravtsov04, Moore06} via supernova
  feedback, photoionization or reionization. This would result in a
  high mass-to-light ratio ($M/L$) in the substructures.  If these
  high-$M/L$ substructures indeed exist, different methods
  for indirect detection are possible. The dark substructure may be
  detectable for example through its effects on stellar streams
  \citep[e.g.,][]{Ibata02, Mayer02}, via $\gamma$-rays from dark
  matter annihilation \citep{Bergstrom99, Calcaneo00, Stoehr03,
  Colafrancesco06} or through gravitational lensing \citep[e.g.,][]{Dalal02,
  Koopmans05}.
    
  \noi While the first two approaches are limited to the local
  Universe, gravitational lensing allows one to explore the mass
  distribution of galaxies outside the Local Group and at a relatively
  high redshift. Moreover, gravitational lensing is independent of the
  baryonic content, of the dynamical state of the system and of the
  nature of dark matter. For example, when in a lens system a point source is close to the caustic fold or cusp, the sum of the image fluxes should add to zero if the sign of the image parities  
  is taken into account \citep{Blandford86,Zakharov95}. This relation is, however, violated by 
  many observed lensed quasars with cusp and
   fold images. 
  As first suggested by \citet{Mao98}, these flux ratio anomalies
  can be related to the presence of (dark matter) substructure around the
  lensing galaxy on scales smaller than the image
  separation \citep{Bradac02, Chiba02, Dalal02,
  Metcalf02, Keeton03, Kochanek04, Bradac04, Keeton05}.
  Nevertheless subsequent studies of similar
  gravitationally lensed systems have shown that
  the required mass fraction in substructure is higher than what is
  obtained in numerical simulations \citep{Mao04, Maccio06,Diemand07b}. In
  addition, for a significant number of cases the observed flux ratio
  anomalies can be explained by taking into account the luminous dwarf
  satellite population \citep{Trotter00, Ros00,
  Koopmans02, Kochanek04, Chen07, McKean07, More08}. Whether the mass fraction
  of CDM substructures is quantifiable via flux ratio anomalies is
  therefore a question still open for debate. Alternatively,
  \citet{Koopmans05} showed that dark matter substructure in lensing
  galaxies can be detected by modelling of multiple images or Einstein
  rings from extended sources. \\
  
  \noi In this paper, we developed an adaptive grid-based modelling
  code for extended lensed sources and grid-based potentials, to fully
  quantify this procedure.  The method presented here is a significant
  improvement of the techniques introduced by \citet{Warren03},
  \citet{Dye05}, \citet{Koopmans05}, \citet{Suyu106},
  \citet{suyu206} and \citet{Brewer06}. In order to detect mass substructure in lens
  galaxies one needs to solve simultaneously for the source surface
  brightness distribution and the lens potential.  A semilinear
  technique for the reconstruction of grid-based sources, given a
  parametric lens potential, was first introduced by
  \citet{Warren03}. The method was subsequently extended by
  \citet{Koopmans05} and  \citet{Suyu106} in order to include a
  grid-based potential for the lens and by \citet{Barnabe07} to
  include galaxy dynamics. \citet{Dye05} introduced an
  adaptive gridding on the source plane; this would minimize the
  covariance between pixels and decrease the computational
  effort. However the method is still lacking an objective procedure
  to quantify the level of regularization. \citet{suyu206} and \citet{Brewer06} encoded the
  semi-linear method within the framework of Bayesian statistics
  \citep{MacKay92, MacKay03}. Although a vast improvement, the fixed
  grids do not allow to take into account the correct number of
  degrees of freedom and proper evidence comparison is difficult.  
  In the implementation here described, these issues have
  been solved:
  
  \smallskip
  
  \noi {\bf (i)} the procedure is fully Bayesian; this allows us to
  determine the best set of non-linear parameters for a given
  potential and the linear parameters of the source, to objectively
  set the level of regularization and to compare/rank different
  potential families;
  
  \smallskip
  
  \noi {\bf (ii)} using a Delaunay tessellation, the source grid
  automatically adaptives in such a way that the computational effort
  is mostly concentrated in high magnification regions;

  \smallskip
  
  \noi {\bf (iii)} the source-grid triangles are re-computed at every
  step of the modelling so that the source and the image plane always
  perfectly map onto each other and the number of degrees of freedom
  remains constant during Bayesian evidence maximisation.
  
  \smallskip
  
  \noi For the first time in the framework of grid-based lensing
  modelling, we use the Nested-Sampling technique by
  \citet{Skilling04} to compute the full marginalized Bayesian
  evidence of the data \citep{MacKay92, MacKay03}.  This approach not
  only provides statistical errors on the lens parameters, but also
  consistently quantifies the relative evidence of a smooth potential
  against one containing substructures.  As such, our method
  provides a fully objective way to rank these two hypotheses given
  the data, which is the goal set out in this paper.
  
  \noi The paper is organized as follow. In Section 2 we give a
  general overview on the data model. In Section 3 we present in
  detail how the data model can be inverted and the source and lens
  potential reconstructed.  In Section 4 we review the basics of
  Bayesian statistics and of the Nested-Sampling technique for
  evidence computation.  In Section 5 we describe how the method has
  been tested and how its ability in detecting substructures,
  depending on the perturbation mass and position, has been
  studied. Finally in Section 6 conclusions are drawn and future
  applications are discussed.
  

  \section{Construction of the lensing operators}
  
  In this section, we describe the data model which relates the
  unknown source brightness distribution and lens potential to the
  known data of the lensed images. The aim is to put this procedure in
  a fully self-consistent mathematical framework, excluding as much as
  possible any subjective intervention into the modelling.  The core
  of the method presented here is based on a Occam's razor argument.
  From a Bayesian evidence point of view, correlated features in the
  lensed images are most likely due to structure in the source, rather
  than being the result of small-scale perturbations of the lens
  potential in front of all the lensed images.  On the other hand,
  uncorrelated structure in the lensed images is most likely due to
  small-scale perturbations of the lens potential.
  
  
  \subsection{The data, source and potential grids}\label{sec:grids}
\begin{figure} 
    \begin{center} 
      \includegraphics[width=\hsize]{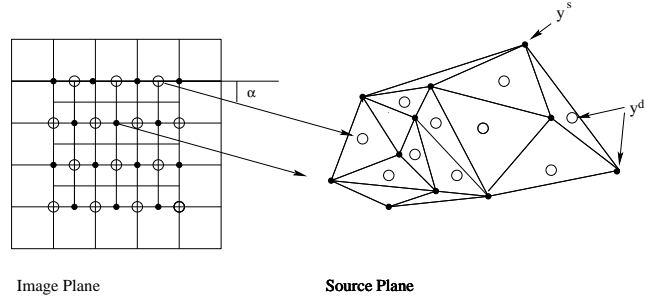}
      \caption {A schematic overview of the non-linear source and
	potential reconstruction method, as implemented in this
	paper. On the left hand-side, on the image plane, two grids
	are defined: one for the potential corrections and one for the
	lensed image. A subset of $N_s$ of the $N_d$ image pixels
	located at the positions $\bmath x^s_i$ on the image plane
	(filled circles) is cast back to the source plane (on the
	right) on $\bmath{y}^s_i$ through the lens equation. These
	form the vertices of an adaptive grid on the source plane. The
	remaining image pixels (open circles) are also cast to the
	source plane to the positions $\bmath{y}_i^d$ (we note that
	this set of points includes $\bmath{y}^s_i$). Because the
	source brightness distribution is conserved, i.e $S(\bmath
	x^d_i)=S(\bmath y^d_i)$, the surface brightness at the empty
	circles is represented by a linear superposition of the
	surface brightness at the three triangle vertices that enclose
	it. Similarly the potential correction at a point
	$\bmath{x}_i^{\delta\psi}$ is given by linear interpolation of
	the potential corrections at the surrounding pixels (large
	rectangular pixels on the image plane). }
      \label{fig:grid} 
    \end{center}
  \end{figure}

  The main idea of grid-based lensing techniques is to use a
  grid-based reconstruction of the source and of the lens potential.
  Here we introduce the general geometry of the problem, explicitly
  shown in Fig. \ref{fig:grid}.  Consider a lensed image $\bmath d$
  of an unknown extended source $\bmath s$. Both $\bmath d$ and
  $\bmath s$ are vectors that describe the surface brightness
  distributions on a set of spatial points $\bmath x_i^d$ and $\bmath
  y_j^s$ in the lens and source plane, respectively
  \citep[e.g.,][]{Warren03,Koopmans05, suyu206}. In general, these are
  related through the lens equation ${\bmath y_i^d} = {\bmath x_i^d} -
  {\bmath \nabla} \psi({\bmath x_i^d})$, where ${\bmath x}_i^d$
  corresponds to the spatial position of the surface brightness in the
  $ith$ element of the vector $\bmath d$, i.e. $d_i$ and $\psi({\bmath x_i^d})$
  is the lensing potential, which is described in more detail in a moment. 
  We note that ${\bmath y}_i^d$ does not necessarily directly correspond to the
  elements $\bmath y_j^s$, $jth$ brightness value
  of the vector $\bmath s$. In our implementation, the grid on the
  source plane is fully adaptive and is directly constructed from a
  subset of the $N_d$ pixels in the image plane, with spatial
  boundaries of the image grid included.  In particular, as shown
  schematically in Fig. \ref{fig:grid}, $N_s$ pixels, located each
  at a position $\bmath x_i^s$ on the image grid, are cast back to the
  source plane giving the positions $\bmath y_j^s$. 
  The set of positions $\{ \bmath y_i^s \}$ constitute
  the vertices of a Delaunay triangulation. In this way, we define an
  irregular adaptive grid, where vertex positions in the source plane
  are related to positions on the image plane via the lens equation
  and every vertex value represents an unknown source surface
  brightness level.  
  
  \noi We assume the lens potential to be the
  superposition of a parametric smooth component with linear local
  perturbations related to the presence of e.g. CDM substructures or
  dwarf galaxies:
  \begin{equation}
    \psi(\bmath x,\bmath \eta)=\psi_s(\bmath x,\bmath
    \eta)+\delta\psi(\bmath x).
  \end{equation}
  While $\psi_s(\bmath x,\bmath \eta)$ assumes a parametric form,
  with parameters $\bmath\eta$, $\delta \psi(\bmath x)$ is a function
  that is pixelized on a regular Cartesian grid of points $\bmath
  x_k^{\delta\psi}$ with values
  $\delta \psi_k$. The set $\{\delta \psi_k\}$ is written as a vector
  $\delta\bmath{\psi}$. Given the observational set of data $\bmath d$,
  we now wish to recover the source distribution $\bmath s$ and the
  lens potential $\psi({\bmath x}, \bmath\eta)$ simultaneously. To do
  this we need to mathematically relate the brightness values $\bmath
  d$ to the unknown brightness values $\bmath s$. As described in the
  next subsection, this can be done through a linear operation on
  $\bmath s$ and $\delta \bmath{\psi}$, where the operator itself is a
  function of an initial guess of the lens potential.

  \subsection{The source and potential operator}

  We now derive the explicit relation between the unknown source
  distribution $\bmath s$, the potential correction $\delta
  \bmath{\psi}$, the smooth potential $\psi_s(\bmath x,\bmath\eta)$
  and the image brightness $\bmath d$. 

\begin{figure}
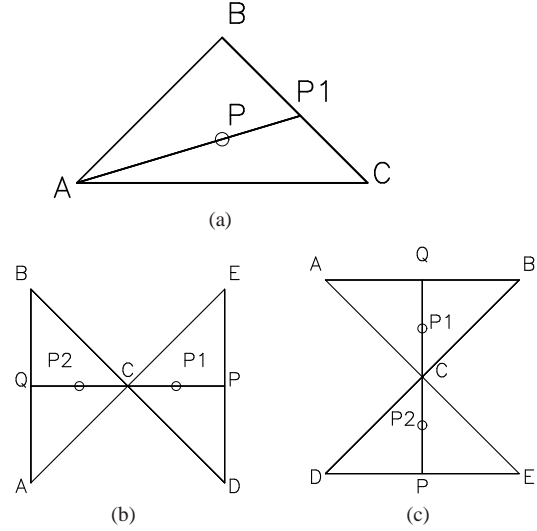

    \begin{center}
      \subfigure[]{\centering \includegraphics[width=4.5cm]{fig2a}
	\label{fig:single}
      }
      \hspace{.5in}

      \subfigure[]{ \includegraphics[width=3cm]{fig2b}
	\label{fig:double_x}
      }
      \hspace{.25in} \subfigure[]{
      \includegraphics[width=3cm]{fig2c}
	\label{fig:double_y}
      }
      
      \caption{Generic triangles from the
	source grid. Both the source surface brightness and its
	derivatives at the points P, $\rm P_1$ and $\rm P_2$ are given
	by linear superposition of the values at the edges of the
	surrounding triangles.}
      \label{fig:triangles} 
      
    \end{center} 
  \end{figure}

  \noi Consider a generic triangle $\widehat{\rm{ABC}}$ on the source
  plane (Fig. \ref{fig:single}), then the source surface brightness
  ${s_{\rm P}}$ on a point P, located inside the triangle at the
  position ${\bmath y}_{\rm P}^d$, can be related to the surface brightness on
  the vertices A, B and C through a simple linear relation 
  \begin{equation}
    {s_{\rm P}}=w_{\rm A}{s_{\rm A}}+w_{\rm B}{s_{\rm B}}+w_{\rm
	C}{s_{\rm C}}\,.
  \end{equation} 
  \noi An explicit expression for the bilinear interpolation weights
  $w_{\rm{A}}$, $w_{\rm B}$ and $w_{\rm C}$ can be obtained by
  considering the point $\rm P_1 $, at the intersection of the line
  $\overline{\rm {AP}}$ with the line $\overline{\rm{CB}}$. The source
  intensities at P and $\rm P_1$ are also related to each other
  through a linear interpolation.  On the other hand, the surface
  brightness in $\rm P_1$ is directly related to the values on the
  triangle vertices $\rm B$ and $\rm C$
  \begin{equation}
    \left\{
    \begin{array}{l}
      s_{\rm P} = \frac{d_{\rm {PA} }
      }{d_{\rm{P_1A}}}(s_{\rm{P_1}}-s_{\rm A})+s_{\rm A}\\ s_{\rm
      {P_1}}= \frac{ d_{ \rm {P_1B} } }{ d_{\rm{CB}} }(s_{\rm
      C}-s_{\rm B})+s_{\rm B}
    \end{array}
    \right.\,
    \label{equ:arr}
  \end{equation}
  \noi where $d_{\rm {PA}}$ and $d_{\rm {P_1A}}$ are the absolute
  distances between the points P and A and the points $\rm P_1$ and A
  respectively; $d_{ \rm{P_1B}}$ and $d_{\rm {CB}}$ are the distances
  between the points $\rm {P_1}$ and B and the points C and B
  respectively. Solving (\ref{equ:arr}), we obtain the weights
  \begin{equation}
    \left\{
    \begin{array}{l}
      w_{\rm A}= 1-\frac{d_{\rm {PA}}}{d_{\rm{P_1A}}}\\ w_{\rm B}=
      \frac{d_{\rm{PA}}}{d_{\rm{P_1A}}}
      \left(1-\frac{d_{\rm{P_1B}}}{d_{\rm{CB}}}\right)\\
      w_{\rm C}=
      \frac{d_{\rm{PA}}d_{\rm{P_1B}}}{d_{\rm{P_1A}}d_{\rm{CB}}}
    \end{array}
    \right.\,
  \end{equation}
  \noi with $\sum_{i=\rm A,\rm B,\rm C }{w_i}=1$. Because
  gravitational lensing conserves the surface brightness, i.e.\ $S(\bmath
  x_i^d) = S(\bmath y_i^d)$, the mapping between the two planes (when
  $\delta\bmath\psi=0$) can be expressed as a system of $N_s$ coupled
  linear equations 
  \begin{equation}
    \mathbf{B\,L}(\bmath \eta)\bmath s =\bmath d + \bmath n\,,
    \label{equ: src_linear_blurred} 
  \end{equation}
  where $\mathbf L(\bmath \eta)$ and $\mathbf B$ are the lensing and
  the blurring operators respectively \citep[see e.g.][]{Warren03,
  Treu04, Koopmans05, Suyu106}. The blurring operator is a square
  sparse matrix which accounts for the effects of the PSF. Each row of
  the lensing operator (a sparse matrix) contains at most the three
  bilinear interpolation weights, $w_{\rm A ,B, C}$, placed at the columns that
  correspond to the three source vertices that enclose the associated
  source position. For a vertex point, there is only one weight equal
  to unity. In case $N_s = N_d$ (i.e.\ all image positions are used to
  create the source grid), all weights are equal to unity. In this
  case, the systems of equations is under-constrained and strong
  regularization is required.

  \noi By pixelating $\delta \psi(\bmath x)$ on a regular Cartesian
  grid, a similar argument as for the source can be applied to the
  potential correction; all potential values, $\{\delta \psi_k\}$, and
  their derivatives on the image plane can be related to this limited
  set of points through bilinear interpolation
  \citep[see][]{Koopmans05, Suyu08}. It is then possible to derive from
  equation~(\ref{equ: src_linear_blurred}) a new set of linear
  equations,
  \begin{equation}
    \mathbf{M_c}\left(\bmath{\eta},\bmath\psi\right)\,\bmath r = \bmath d +
    \bmath n,
    \label{equ: src_pot_linear_blurred}
  \end{equation} 
  where
  \begin{equation}
    \bmath r\equiv\left(
    \begin{array}{c}
      \bmath s\\ \delta\bmath \psi
    \end{array} 
    \right)\,.
  \end{equation}
  \noi More specifically, $\bmath\psi$ is the sum of all the previous
  corrections $\delta\bmath\psi$ and the operator $\mathbf{M_c}$ is a
  block matrix reading
  \begin{equation}
    \mathbf{M_c}\equiv \mathbf B \left [\mathbf L(\bmath \eta, \bmath
      \psi)\, | -\mathbf{D_s}(\bmath s_{\rm MP})\mathbf {D_{\psi}}\\
      \right]\,.
    \label{equ:block_matrix}
  \end{equation} 
  \noi ${\mathbf L}({\bmath \eta}, {\bmath \psi})$ is the
  lensing operator introduced above, $\mathbf{D_s}(\bmath s_{\rm MP})$
  is a sparse matrix whose entries depend on the surface brightness
  gradient of the previously-best source model at $\bmath{y}^d_i$ and
  $\mathbf{D_\psi}$ is a matrix that determines the gradient of
  $\delta\bmath\psi$ at all corresponding points $\bmath{x}^d_i$
  \citep[see] [for details]{Koopmans05}. The generic structure of
  these matrices is given by
  \begin{equation}
    \mathbf{D_{s}}= \left(
    \begin{array}{ccccc}
      ...&&& \\ \\ &\frac{\partial S({\bmath y}^d_i)}{\partial y_1}&
      \frac{\partial S({\bmath y}^d_i)}{\partial y_2} & \\ \\ &
      &\frac{\partial S({\bmath y}^d_{i+1})}{\partial y_1}&
      \frac{\partial S({\bmath y}^d_{i+1})}{\partial y_2} \\ \\ & & & & ...\\
    \end{array}
    \right)
  \end{equation}
  and
  \begin{equation}
    \mathbf{D_{\delta\psi}}= \left(
    \begin{array}{ccccc}
      ...&\\ \\ &\frac{\partial \delta\psi (\bmath{x}^d_i)}{\partial
      x_1}&\\ &\frac{\partial \delta\psi (\bmath{x}^d_i)}{\partial
      x_2} & \\ \\ &&\frac{\partial \delta\psi (\bmath{x}^d_{i+1})}{\partial
      x_1} \\ &&\frac{\partial \delta\psi (\bmath{x}^d_{i+1})}{\partial
      x_2} \\ & & & &...\\
    \end{array}
    \right)
  \end{equation}
  where the index $i$ runs along all the $\bmath{x}_i^d$ and $\bmath{y}_i^d$,
  i.e. triangle vertices included. The ``functions'' $S$ and $\delta
  \psi$ and their derivative can be derived through bilinear
  interpolation and finite differencing from $\bmath s$ and $\delta
  \bmath \psi$, respectively.

  \noi It is clear from the structure of these matrices that the
  first-order correction to the model, as a result of $\delta \psi$,
  is equal to $\delta d_i= -\bmath {\nabla} S(\bmath{y}^d_i) \cdot
  \bmath{\nabla} \delta \psi(\bmath{x}^d_i)$ at every point
  $\bmath{x}^d_i$ \cite[see e.g.][for a derivation]{Koopmans05}.

  \noi As for the surface brightness itself, also the first derivatives for
  a generic point P on the source plane can be expressed as functions
  of the relative values on the triangle vertices A, B, C, yielding
  \begin{eqnarray}
    \frac{\partial {s_{\rm P}}}{\partial y_{1}} & = &w_{\rm
      A}\frac{\partial {s_{\rm A}}}{\partial y_{1}}+w_{\rm
      B}\frac{\partial {s_{\rm B}}}{\partial y_{1}}+w_{\rm
      C}\frac{\partial {s_{\rm C}}}{\partial y_{1}}\nonumber\\
      \frac{\partial {s_{\rm P}}}{\partial y_{2}} & = &w_{\rm
      A}\frac{\partial {s_{\rm A}}}{\partial y_{2}}+w_{\rm
      B}\frac{\partial {s_{\rm B}}}{\partial y_{2}}+w_{\rm
      C}\frac{\partial {s_{\rm C}}}{\partial y_{2}}
  \end{eqnarray}
  For the generic vertex $j= \rm{A, B,C}$ these are given by
  $\frac{\partial \bmath{s_j}}{\partial y_{1}}=-\frac{n_0}{n_2}$
  and $\frac{\partial \bmath{s_j}}{\partial
  y_{2}}=-\frac{n_1}{n_2}$, where  $\bmath{N}\equiv(n_0,n_1,n_2)$ is the
  unit-length surface normal vector at the vertex $j$ and is defined
  as the average of the adjacent per-face normal vectors. For
  $\delta\bmath\psi$ and its gradients, on a rectangular grid with
  rectangular pixels, we follow \cite{Koopmans05}.
  
  \section{Inverting the data model}\label{sec:inverting}

  \noi As shown above, in both cases of solving for the source alone,
  or solving for the source plus a potential correction, a {\sl linear
  data model} can be constructed. In this section, we give a
  general overview of how this set of linear equations can be
  (iteratively) solved. A more thorough Bayesian description and
  motivation can be found in Section~4.
  
  \subsection{The penalty function}
  Before we go into the details of the method, we first restate that
  for a given lens potential $\psi(\bmath x, {\bmath \eta})$ and
  potential correction $\bmath \psi_n = \sum^n_{i=1}
  \delta {\bmath \psi_i}$, on a grid, the source surface brightness vector
  $\bmath s$ and the data vector $\bmath d$ can be related through a
  linear (matrix) operator
  \begin{equation}
    \mathbf {M_c}({\bmath \eta}, {\bmath \psi}_{n-1}, \bmath
    s_{n-1})\bmath r_n={\bmath d} + {\bmath n},
    \label{equ: src_linear} 
  \end{equation}
  now explicitly written with their dependencies on the source and
  potential and with
  \begin{equation}
    \bmath r_n= \left(\begin{array}{c}\bmath s_{n} \\ 
      \delta\bmath\psi_n \\
    \end{array}
    \right).
  \end{equation}
  In this equation $\bmath s_n$ is a model of the source
  brightness distribution at a given iteration $n$ (we describe the
  iterative scheme momentarily). We assume the noise $\bmath n$ to be
  Gaussian which is a good approximation for the HST images the 
  method will be applied to. Even in case of deviations from Gaussianity, 
  the central limit theorem, for many data points, ensures that the probability density 
  distribution is often well approximated by a Normal distribution. \\
  \noi Because of the ill-posed nature of this relation,
  equation (\ref{equ: src_linear}) cannot simply be inverted. Instead a
  penalty function which expresses the mismatch between the data and
  the model has to be defined by
  \begin{equation}\label{eqn:penalty}
    P(\bmath s,\delta \bmath \psi \,|\, {\bmath \eta}, {\bmath \lambda},
    {\bmath s}_{n-1}, {\bmath
    \psi}_{n-1})=\chi^2+\lambda_s^2\|\mathbf{H_s} \bmath s\|^2_2
    +\lambda_{\delta\psi}^2 \|\mathbf{H_{\delta\psi}} \delta\bmath
    \psi\|^2_2\,,
  \end{equation}
  with
  \begin{equation}\label{eqn:chi2}
    \chi^2 = [\mathbf {M_c}({\bmath \eta}, \bmath \psi_{n-1}, \bmath
    s_{n-1})\, \bmath r - {\bmath d}]^{\rm T} \, {\mathbf {C_d^{-1}}} \,
    [\mathbf {M_c}({\bmath \eta}, \bmath \psi_{n-1}, \bmath
    s_{n-1})\,\bmath r - {\bmath d}].
  \end{equation}
  
  \noi The second and third term in the penalty function contain prior
  information, or beliefs about the smoothness of the source and of
  the potential respectively and $\mathbf{C_d}$ is the diagonal
  covariance matrix of the data. The level of regularization is set by
  the regularization parameters $\bmath \lambda$, one for the source and one
  for the potential \citep[see][for a more general
  discussion]{Koopmans05, suyu206}.  In a Bayesian framework, this
  penalty function is related to the posterior probability of the
  model given the data (see Section 4). In the following two sections
  we describe how to solve for the linear and non-linear parameters of
  the penalty function (except for $\bmath \lambda$, which is described
  in Section 4).
  
  \subsubsection{Solving for the linear parameters}
  \label{sec:solvelinear}
  The most probable solution, $\bmath{r_{\rm MP}}$, minimizing the
  penalty function is obtained by solving the set of linear equations
  \begin{equation}
    (\mathbf{M_c^T C_d^{-1}M_c+R^T R})\,\bmath
    r=\mathbf{M_c^TC_d^{-1}}\bmath d.
    \label{equ: src_pot_penalty} 
  \end{equation}
  The regularization matrix is given by
  \begin{equation}
    {\mathbf R^{\rm T}} {\mathbf R} = \left(
    \begin{array}{cc}
      \lambda_s^2\mathbf{H_s^{\rm T}} \mathbf{H_s} & \\ &
      \lambda^2_{\delta\psi}\mathbf{H_{\delta\psi}^{\rm T}}
      \mathbf{H_{\delta\psi}}
    \end{array} \right).
  \end{equation}
  
  \noi The solution of this symmetric positive definite set of
  equations can be found using e.g.\ a Cholesky decomposition
  technique. By solving equation (\ref{equ: src_pot_penalty}), adding
  the correction $\delta \bmath \psi_n$ to the previously-best
  potential $\bmath \psi_{n-1}$ and iterating this procedure, both the
  source and the potential should converge to the minimum of the
  penalty function $P(\bmath s_n,\delta \bmath \psi_{n} \,|\, {\bmath
  \eta}, {\bmath \lambda}, {\bmath s}_{n-1}, {\bmath \psi}_{n-1})$. At
  every step of this iterative procedure the matrices $\mathbf {M_c}$
  and $\mathbf R$ have to be recalculated for the new updated
  potential $\bmath \psi_n$ and source $\bmath s_n$. While the
  potential grid points are kept spatially fixed in the image plane,
  the Delaunay tessellation grid of the source is re-built at every
  iteration to ensure that the number of degrees of freedom is kept
  constant during the entire optimization process.
  
  \noi Note that because the source and the potential corrections are
  independent, they require their own form ($\mathbf H$) and level
  ($\lambda$) of regularization.  The most common forms of
  regularization are the zeroth-order, the gradient and the
  curvature. As shown by \citet{suyu206} the best form depends on the
  nature of the source distribution and can be assessed via Bayesian
  evidence maximisation. For the source, we chose the curvature
  regularization defined for the Delaunay tessellation of the source
  plane. 

  \noi Specifically one can combine the gradient and curvature
  matrices in the $x$ and $y$ directions: $\mathbf{H_{s}^{\rm
  T}}\mathbf{H_{s}}=\mathbf{H_{s,y_1}^{\rm
  T}}\mathbf{H_{s,y_1}}+\mathbf{H_{s,y_2}^{\rm T}}\mathbf{H_{s,y_2}}$.
  Both $\mathbf{H_{s,y_1}}$ and $\mathbf{H_{s,y_2}}$ can be obtained
  by analogy by considering the pair of triangles in
  Fig.~\ref{fig:double_x} and Fig.~\ref{fig:double_y}
  respectively.

  \noi For every generic point C on the source plane we consider the
  pair of triangles $\widehat{\rm{ABC}}$ and $\widehat{\rm{DCE}}$ and
  define the curvature in C in the $y_1$ direction as:
  \begin{equation}
    {s''_{C,y_1}}
    \equiv \frac{1}{d_{CP}}({s_P}-{s_C}) -\frac{1}{d_{CQ}}({s_C}-{s_Q})\,.
    \label{equ:curvature}
  \end{equation}
  This is not the second derivative, but we find that this alternative
  curvature definition gives much better results than using the second
  derivative directly. The reason is that it gives equal weight to all
  triangles, independently of their relative sizes (for identical
  rectangular pixels this problem does not arise since the above
  definition is equal to the second derivative up to a proportionality
  constant). A much smoother solution in that case is obtained.
  
  \noi P and Q
   are given by intersecting the line
  $\overline{\rm{CP_1}}$ with the line $\overline{\rm{ED}}$ and the
  line $\overline{\rm{CP_2}}$ with the line $\overline{\rm{AB}}$
  respectively. Specifically, $\rm{P_1}$ and $\rm{P_2}$ are defined as
  very small displacements from the point C in the $y_1$ direction %
  \begin{eqnarray}
    y_{2}^{\rm{P_1}}      & = & y_{2}^{\rm{P_2}} =  y_{2}^{\rm C}\nonumber\\
    y_{1}^{\rm{P_{1,2}}}  & = & y_{1}^{\rm C}  \pm \delta y_1.
  \end{eqnarray}
  The source surface brightness in P and Q can be obtained by
  linear interpolation between the source values in D with the value
  in E and the value in A with the value in B respectively
  \begin{eqnarray}
    s_{\rm P}&=&\frac{d_{\rm{PD}}}{d_{\rm{ED}}}(s_{\rm E}-s_{\rm
      D})+s_{\rm D}\label{equ:s_p} \nonumber \\ s_{\rm
      Q}&=&\frac{d_{\rm{QA}}}{d_{\rm{AB}}}({s_{\rm B}}-s_{\rm
      A})+s_{\rm A}\label{equ:s_q}\,,
  \end{eqnarray}
  \noi Substituting (\ref{equ:s_p}) in
  (\ref{equ:curvature}) gives
  \begin{multline}
    {s''_{C,y_1}}=-\left(\frac{1}{d_{\rm
      {CP}}}+\frac{1}{d_{\rm {CQ}}}\right){s_{\rm C}}+\frac{d_{\rm
      PD}}{d_{\rm CP}d_{\rm DE}}s_{\rm E}+\\ \frac{d_{\rm
      {QA}}}{d_{\rm{CQ}}d_{\rm{AB}}}s_{\rm B}+\frac{d_{\rm{PE}}}
      {d_{\rm{CP}}{d_{\rm{DE}} }}s_{\rm D}+\frac{d_{\rm
      {QB}}}{d_{\rm{CQ}}d_{\rm{AB}}}s_{\rm A}\,.
  \end{multline}
  \noi Each row of the regularization matrix $\mathbf{H_{s,y_1}}$, corresponding to every
  point C, contains the five interpolation weights, placed at the
  columns that correspond to the five vertices A, B, C, D and
  E. The curvature in the $y_2$ direction is derived in an analogous
  way using the pair of triangles in Fig. \ref{fig:double_y}. We
  refer again to \citet{Koopmans05} for details on the
  potential regularization matrix $\mathbf{ H_{\delta \psi}}$
  
  \subsubsection{Solving for the non-linear parameters}
  \label{sec:solvenonlinear}
  In order to recover the non-linear parameters $\bmath \eta$, we need
  to minimize the penalty function $P(\bmath s, {\bmath \eta}\,|\,
  {\bmath \lambda}, {\bmath \psi})$. We allow for a correction,
  $\bmath \psi$, to the parametric potential $\psi(\bmath \eta,\bmath
  x)$ (not necessarily zero), but do not allow it to be changed while
  optimising for $\bmath s$ and ${\bmath \eta}$. In all cases, we keep
  $\bmath \lambda$ fixed during the optimization. Given an
  initial guess for the non-linear parameters $\bmath \eta_0$, we then
  minimize the penalty function defined in Section
  \ref{sec:solvelinear}, under the conditions outlined above
  ($\bmath\psi$ is constant and $\delta\bmath\psi \equiv \bmath 0$).
  We use a non-linear optimizer \citep[in our case Downhill-Simplex
  with Simulated Annealing;][]{Press92}, to change $\bmath \eta$ at
  every step and to minimize the joint penalty function $P(\bmath s,
  {\bmath \eta}\,|\, {\bmath \lambda}, {\bmath \psi})$.  The
  optimization of $\bmath s$ is implicitly embedded in the
  optimization of $\bmath \eta$ by solving equation (\ref{equ:
  src_pot_penalty}) only for $\bmath s$, every time $\bmath \eta$ is
  modified.
  
  \subsection{The optimization strategy}\label{sec:strategy}
  
  We have implemented a multi-fold optimization scheme for solving the
  linear equation (\ref{equ: src_linear}). This scheme is not unique,
  but stabilises the numerical optimization of this rather complex set
  of equations. Solving all parameters simultaneously would be
  computationally prohibitive and usually shows poor convergence
  properties.

  \subsubsection{Optimization steps}
  
  Our optimization scheme is similar to a {\sl line-search}
  optimization, where consecutively different sets of unknown
  parameters are being kept fixed, while the others are optimized
  for. The sets $\{\delta \bmath \psi, \bmath s\}$, $\{\bmath \eta,
  \bmath s \}$ and $\{\bmath \lambda, \bmath s \}$ define the three
  different groups of parameters, of which only one is solved for at
  once. The individual steps, in no particular order, are then:
  
  \noi {\bf (i)} {We assume $\bmath \eta$
    and $\bmath \lambda$ to be constant vectors and iteratively solve
    for $\delta\bmath\psi$ and the source $\bmath s$. In this case, at
    every iteration we solve for $\bmath r$ and adjust $\bmath \psi$,
    using the linear correction to the potential $\delta \bmath
    \psi$. This was described in Section \ref{sec:solvelinear}.}
  
  \noi {\bf (ii)} {We assume $\bmath\psi$ and
    $\bmath \lambda$ to be constant vectors and
    $\delta\bmath\psi_i=\bmath 0$ at every iteration and only solve
    for the non-linear potential parameters $\bmath \eta$ and the
    source $\bmath s$. This was described in Section
    \ref{sec:solvenonlinear}. We note that part of step (i) is also
    implicitly carried out in step (ii) (i.e.\ solving for $\bmath s$).}
  
  \noi {\bf (iii)} {We assume both (i) and (ii), above, and solve for
    the regularization parameters $\lambda_s$ of the source and the source
    itself $\bmath s$. This requires a Bayesian approach and will be
    described in more detail in Section~4. We have not attempted to 
    optimize for $\lambda_{\delta \psi}$, but will study this
    in future publications.}

  \noi The overall goal, however, remains to solve for the \emph{full}
  set of unknown parameters $\{ {\bmath \eta}, {\bmath \psi}_n, \bmath
  s_n \}$ for $n\rightarrow \infty$ (or some large number).  In
  particular if an overall smooth (on scales of the image separations)
  potential model $\psi(\bmath \eta)$ does not allow a proper
  reconstruction of the lens system, we add an additional and more
  flexible potential correction $\delta{\bmath \psi}$,
  which can describe a more complex mass structure. 

  \subsubsection{Line-search optimization scheme}

  In practice, we find that the optimal strategy to minimize the
  penalty function is the following, in order:
  
  \noi {\bf (1)} {We set $\lambda_{\rm s}$ to a large constant value
    such that the source model remains relatively smooth throughout
    the optimization (i.e.\ the peak brightness of the model is a
    factor of a few below that of the data) and keep
    $\bmath\psi_n=\bmath 0$ \citep[see also][]{suyu206, Suyu08}.  We then
    solve for $\bmath \eta$ and $\bmath s$ that minimize the penalty
    function}.
  
  \noi {\bf (2)} {Once the best $\bmath \eta$ and $\bmath s$ are
    found, a Bayesian approach is used to find the best value of
    $\lambda_{\rm s}$ for the source only.  At this point
    $\bmath\psi$ is still kept equal to zero.}
  
  \noi {\bf (3)} {Given the new value of $\lambda_{\rm s}$, step (1) is repeated
    to find improved values of $\bmath \eta$ and $\bmath s$. Since the
    sensitivity of $\lambda_{\rm s}$ to changes in $\bmath \eta$ is
    rather weak, at this point the best values of $\bmath \eta$,
    $\bmath s$ and $\bmath \lambda$ have been found.}
  
  \noi {\bf (4)} {Next, all the above parameters are kept fixed and we
    solve for $\bmath r$, this time assuming a very large value for
    $\lambda_{\delta \psi}$ to keep the potential correction (and
    convergence) smooth. We adjust $\bmath \psi$ at every iteration
    until convergence is reached
    \cite[e.g.][]{Suyu08}. At this point we stop the optimization
    procedure.}
  
  \noi {\bf (5)} {The smooth model with $\bmath \psi = \bmath 0$ and
    the same model with $\bmath \psi \neq \bmath 0$ are then compared
    through their Bayesian evidence values and errors on the
    parameters are estimated through the Nested Sampling of
    \citet{Skilling04}(Section 4).}  

\begin{figure*} 
    \begin{center} 
      \includegraphics[width=\hsize,clip=]{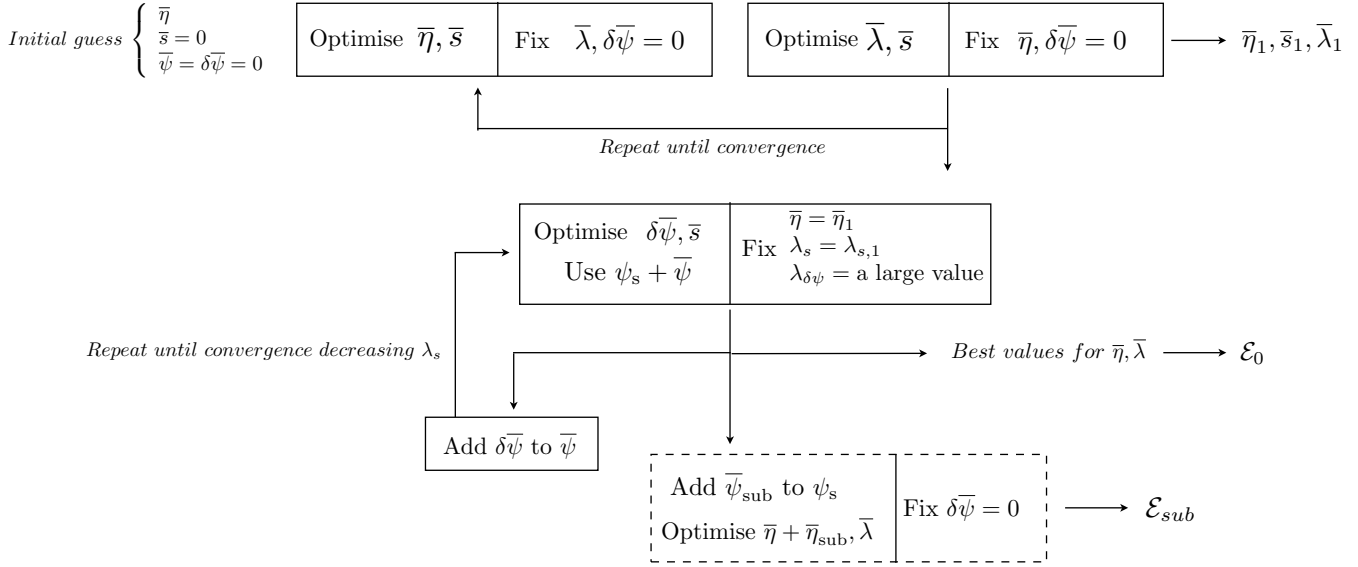}     
       \caption {A schematic overview of the non-linear source and
	potential reconstruction method.}
      \label{fig:flow} 
    \end{center}
  \end{figure*}

  \noi Fig. \ref{fig:flow} shows a complete flow diagram of our
    optimization scheme. In the next section we place
    equation (\ref{eqn:penalty}) and model ranking on a formal Bayesian
    footing. Those readers mostly interested in the application and
    tests of the method could continue reading in Section~5.
     
  \section{A Bayesian approach to data fitting and model selection}
  \label{sec:bayes}
  
  When trying to constrain the physical properties of the lens galaxy,
  within the grid-based approach, three different problems are
  faced.  Given the linear relation in equation (\ref{equ:
  src_pot_linear_blurred}) we need to determine the linear parameters
  $\bmath r$ for a certain set of data $\bmath d$ and a form for the
  smooth potential $\psi_{s}(\bmath x,\bmath \eta)$. We then aim to
  find the best values for the parameters $\bmath \eta$ and $\bmath
  \lambda$ and finally, on a more general level, we wish to infer the
  best model for the overall potential and quantitatively rank
  different potential families. In particular, we want to compare smooth models with models
  that also include a potential grid for substructure (with more free
  parameters). These issues can all be quantitatively and objectively
  addressed within the framework of Bayesian statistics. In the
  context of data modelling three levels of inference can be
  distinguished \citep{MacKay92, suyu206}.
  
  \medskip
  
  \noi {\bf (1)} First level of inference: linear optimization.  We
  assume the model $\mathbf{M_c}$, which depends on a given potential
  and source model, to be true and for a fixed form $\mathbf R$ and
  level ($\bmath\lambda$) of regularization, we derive from Bayes'
  theorem the following expression:
  \begin{equation}
    P\left(\bmath r\,|\,\bmath d,\bmath\lambda,\bmath \eta,\mathbf
    {M_c},\mathbf R\right)=\frac{P(\bmath d \,|\,\bmath r,\bmath \eta,
    \mathbf{M_c})\, P(\bmath r\,|\,\bmath\lambda,\mathbf R)}{P(\bmath
    d \,|\,\bmath\lambda,\bmath \eta,\mathbf{M_c},\mathbf R)}\,.
  \end{equation}
  The likelihood term, in case of Gaussian noise, for a covariance
  matrix $\mathbf{C_d}$, is given by
  \begin{equation}
    P(\bmath d \,|\,\bmath r, \bmath\eta,\mathbf{M_c})=
	\frac{1}{Z_d}\exp{[-E_d(\bmath d \,|\,\bmath
	r,\bmath\eta,\mathbf{M_c})]}\,
  \end{equation}
  where
  \begin{equation}
    Z_d=(2\pi)^{N_d/2}(\det \ \mathbf{C_d})^{1/2}
  \end{equation}
  and (see equation \ref{eqn:chi2})
  \begin{equation}
    E_d(\bmath d \,|\,\bmath r,\bmath\eta,\mathbf{M_c}]=
      \frac{1}{2}\,\chi^2=\frac{1}{2}\left(\mathbf{M_c} \bmath
      r-\bmath d\right)^{\rm T}\mathbf{C}_D^{-1}\left(\mathbf{M_c}
      \bmath r-\bmath d\right)\,.
  \end{equation}
  Because of the presence of noise and often the singularity of
  $\det\,(\mathbf{M_c^{\rm T}} \mathbf{M_c})$, it is not possible to
  simply invert the linear relation in equation (\ref{equ:
  src_pot_linear_blurred}) but an additional penalty function must be
  defined through the introduction of a prior probability $P(\bmath r
  \,|\,\bmath\lambda,\mathbf R)$ on $\bmath s$ and on $\delta\bmath
  \psi$. In our implementation of the method, the prior assumes a
  quadratic form, with minimum in $\bmath r=\bmath 0$ and sets the
  level of smoothness (specified in $\mathbf H$ and $\bmath\lambda$)
  for the solution
  \begin{equation}
    P(\bmath r\,|\,\bmath\lambda,\mathbf R)=
    \frac{1}{Z_r}\exp{\left[-\bmath\lambda E_r(\bmath r\,|\,\mathbf
    R)\right]}\,,
  \end{equation}
  with
  \begin{equation}
    Z_r(\bmath\lambda)=\int {d\bmath r e^{-\bmath\lambda E_r}}=
    e^{-\bmath\lambda
    E_s(0)}\left(\frac{2\pi}{\bmath\lambda}\right)^{N_r/2}(\det\mathbf
    C)^{-1/2}\,,
  \end{equation}
  \begin{equation}
    E_r=\frac{1}{2}\|\mathbf R\bmath r\|^2_2
  \end{equation}
  and
  \begin{equation}
    \mathbf C=\nabla \nabla E_r=\mathbf R\,\mathbf {R}^{\rm T}\,.
  \end{equation}
  The normalization constant $P(\bmath d\,|\,\bmath\lambda,\bmath
  \eta,\mathbf{M_c},\mathbf R)$ is called the evidence and plays an
  important role at higher levels of inference. In this specific case
  it reads
  \begin{equation}
    P(\bmath d\,|\,\bmath\lambda,\bmath \eta,\mathbf{M_c},\mathbf R)
    =\frac{\int{d\bmath r\exp{(-M(\bmath r))}}}{Z_d Z_r}\,,
  \end{equation}
  \noi where
  \begin{equation}
    M(\bmath r)=E_d+ E_r\,.
  \end{equation}
  The most probable solution for the linear parameters, is found by
  maximizing the posterior probability
  \begin{equation}
    P(\bmath r\,|\,\bmath d,\bmath\lambda,\bmath
    \eta,\mathbf{M_c},\mathbf R)=\frac{\exp(-M(\bmath
    r))}{\int{d\bmath r\,\exp(-M(\bmath r))}}\,.
    \label{equ:posterior}
  \end{equation}
  The condition $\partial (E_d+ E_r)/\partial \bmath r=0$ now yields the
  set of linear equations already introduced in Section
  \ref{sec:solvelinear}:
  \begin{equation}
    \left(\mathbf{M_c^{\rm T}} \mathbf{C_d}^{-1} \mathbf{M_c}+\mathbf
    R^{\rm T} \mathbf R\right)\bmath r = \mathbf{M_c^{\rm T}}
    \mathbf{C_d}^{-1}\bmath d\,.
    \label{equ:src_pot_penalty_bayes}
  \end{equation}
  Equation (\ref{equ:src_pot_penalty_bayes}) is solved iteratively
  using a Cholesky decomposition technique.  
  
  \noi {\bf (2)} Second level of inference: non-linear optimization.
  At this level we want to infer the non-linear parameters $\bmath
  \eta$ and the hyper-parameter $\lambda_{\rm s}$ for the
  source. Since at this point we are interested only in the smooth
  component of the lens potential, we set $\delta\bmath \psi=0$ and
  for a fixed family $\psi_s(\bmath \eta)$, form of the regularization
  $\mathbf R$ and model $\mathbf{M_c}$, we maximize the posterior
  probability
  
  \begin{equation}\label{equ:posterior_2}
    P(\bmath\lambda,\bmath \eta\,|\,\bmath d,\mathbf{M_c},\mathbf
      R)=\frac{P(\bmath d\,|\,\bmath \lambda,\bmath \eta,\mathbf{M_c},\mathbf
      R)P(\bmath \lambda,\bmath \eta)}{P(\bmath d\,|\,\mathbf{M_c},\mathbf
      R)}\,.
  \end{equation}
  
  \noi Assuming a prior $P(\bmath \lambda,\bmath \eta)$, which is flat in
  $\log(\lambda_s)$ and $\bmath\eta$, reduces to maximizing the
  evidence $P(\bmath d\,|\,\bmath\lambda,\bmath
  \eta,\mathbf{M_c},\mathbf R)$ (which here plays the role of the
  likelihood) for $\bmath \eta$ and $\bmath\lambda$. The evidence can
  be computed by integrating over the posterior (\ref{equ:posterior_2})
  \begin{equation}
    P(\bmath d\,|\,\bmath\lambda,\bmath \eta,\mathbf{M_c},\mathbf R)=\int{d\bmath
      r\, P(\bmath d\,|\,\bmath r,\bmath
      \eta,\mathbf{M_c})P(\bmath r\,|\,\bmath\lambda,\mathbf
      R)}\,.
    \label{equ:evidence}
  \end{equation}
  Because of the assumptions we made (Gaussian noise and quadratic
  form of regularization), this integral can be solved analytically
  and yields
  \begin{equation}
    P(\bmath d\,|\,\bmath\lambda,\bmath \eta,\mathbf{M_c},\mathbf R)=
    \frac{Z_M(\bmath\lambda, \bmath \eta)}{Z_d Z_r(\bmath\lambda)}\,,
  \end{equation}
  where
  \begin{equation}
    Z_M(\bmath\lambda, \bmath \eta)=\exp{(-M(\bmath
      r_{\rm MP}))}\left(2\pi\right)^{N_r/2}(\det \ \mathbf A)^{-1/2}\,,
  \end{equation}

 \noi  with $\mathbf A=\nabla\nabla M(\bmath r).$ Again we proceed in an
  iterative fashion: using a simulated annealing technique we maximize
  the evidence (\ref{equ:evidence}) for the parameters $\bmath
  \eta$. Every step of the maximisation generates a new model
  $\mathbf{M_c}(\psi(\bmath \eta_i))$, for which the most probable
  source $\bmath s_{\rm{MP}}$ is reconstructed as described in Section
  \ref{sec:inverting}. At this starting step the level of the source
  regularization is set to a relatively large initial value
  $\lambda_{s,0}$; in this way we ensure the solution to be smooth (at
  least at this first level) and the exploration of the $\bmath \eta$
  space to be faster. Subsequently we fix the best model
  $\mathbf{M_c}(\bmath \eta_0)$ found at the previous iteration and,
  using the same technique, we maximize the evidence for the source
  regularization level $\lambda_s$.  The procedure is repeated until
  the total evidence has reached its maximum. In principle we should
  have built a nested loop for $\lambda_s$ at every step of the
  $\bmath \eta$ exploration, but in practice the regularization
  constant only changes slightly with $\bmath \eta$ and the alternate
  loop described above gives a faster way to reach the maximum
  (line-search method).
  
  \noi {\bf (3)} At the third level of inference Bayesian statistics
  provides an objective and quantitative procedure for model
  comparison and ranking on the basis of the evidence,
  \begin{equation}
    P(\mathbf{M_c},\mathbf R\,|\,\bmath d) \propto P(\bmath
    d\,|\,\mathbf{M_c},\mathbf R)P(\mathbf{M_c},\mathbf R)\,.
  \end{equation}
  For a flat prior $P(\mathbf{M_c},\mathbf R)$ (at this level of
  inference we can make little to no assumptions) different models can
  be compared according to their value of $P(\bmath
  d\,|\,\mathbf{M_c},\mathbf R)$, which is related to the evidence of
  the previous level by the following relation
  \begin{equation}
    P(\bmath d\,|\,\mathbf{M_c},\mathbf R)=\int{d\bmath\lambda\, d\bmath
      \eta \,P(\bmath d\,|\,\bmath \lambda,\bmath \eta,\mathbf{M_c},\mathbf
      R) P(\bmath\lambda,\bmath\eta)}\,.
    \label{equ:evidence_integral}
  \end{equation}
  Being multidimensional and highly non-linear, the integral
  (\ref{equ:evidence_integral}) is carried out numerically through a
  Nested-Sampling technique \citep{Skilling04}, which is described in
  more detail in the next section. A by-product of this method is an
  exploration of the posterior probability (\ref{equ:posterior_2}),
  allowing for error analysis of the non-linear parameters and of the
  evidence itself.
  
  \subsection{Model selection: smooth versus clumpy models}\label{sec:nested sampling} 
  
  In the previous section we introduced the main structure of the
  Bayesian inference for model fitting and model selection. While
  parameter fitting simply determines how well a model matches the
  data and can be easily attained with the relatively simple analytic
  integrations of the first and second level of inference, model
  selection itself requires the highly non-linear and multidimensional
  integral (\ref{equ:evidence_integral}) to be solved.  This
  marginalized evidence can be used to assign probabilities to models
  and to reasonably establish whether the data require or allows
  additional parameters or not. Given two competing models $\rm M_0$
  and $\rm M_1$ with relative marginalized evidence ${\cal{E}}_0$ and
  ${\cal{E}}_1$, the Bayes factor, $\Delta {\cal{E}} \equiv
  \log{\cal{E}}_0 - \log{\cal{E}}_1$, quantifies how well $\rm M_0$ is
  supported by the data when compared with $\rm M_1$ and it
  automatically includes the Occam's razor. Typically the literature
  suggests to weigh the Bayes factor using  Jeffreys' scale
  \citep{Jeffreys61}, which however provides only a qualitative
  indication: $\Delta {\cal{E}} < 1$ is not significant, $1 < \Delta
  {\cal{E}}< 2.5$ is significant, $2.5 < \Delta {\cal{E}}< 5$ is
  strong and $\Delta {\cal{E}} > 5$ is decisive.

  \noi In order to evaluate this marginalized evidence with a high
  enough accuracy we implemented the new evidence algorithm known as
  Nested Sampling, proposed by \citet{Skilling04}. Specifically, we
  would like to compare two different models: one in which the lens
  potential is smooth and one in which substructures are present, with
  e.g. a NFW profile. While the first is defined by the non-linear
  parameters of the lens potential and of the source regularization
  only, the second also allows for three extra parameters: the mass of
  the substructure and its position on the lens plane (see
  Section \ref{sec:test})
  
  \subsection{Model ranking: nested sampling}
  
  Here, we provide a short description of how the Nested Sampling can
  be used to compute the marginalized evidence and errors on the model
  parameters; a more detailed one can be found in
  \citet{Skilling04}. The Nested-Sampling algorithm integrates the
  likelihood over the prior volume by moving through thin nested
  likelihood surfaces. Introducing the fraction of total prior
  mass $X$, within which the likelihood exceeds ${\cal L^*}$, hence
  \begin{equation}
    X=\int_{{\cal{L}}>{\cal{L^*}}}{dX}\,,
  \end{equation}
  with
  \begin{equation}
    dX=P\left(\bmath\lambda,\bmath\eta\right)d\bmath\lambda\,d\bmath\eta\,,
  \end{equation}
  the multi-dimensional integral (\ref{equ:evidence_integral})
  relating the likelihood $\cal{L}$ and the marginalized evidence
  $\cal{E}$ can be reduced to a one-dimensional integral with positive
  and decreasing integrand
  \begin{equation}
    {\cal{E}}=\int_0^1{dX\,{\cal{L}}(X)}\,.
  \end{equation}
  
  \noi Where ${\cal L}(X)$ is the likelihood of the (possibly disjoint)
  iso-likelihood surface in parameter space which encloses a total prior
  mass of $X$. If the likelihood ${\cal{L}}_j={\cal{L}}(X_j)$ can be
  evaluated for each of a given set of decreasing points, $0 < X_j <
  X_{j-1} <....< 1$, then the total evidence ${\cal{E}}$ can be
  obtained, for example, with the trapezoid rule,
  ${\cal{E}}=\sum_{j=1}^m{\cal{E}}_j=\sum_{j=1}^m{\frac{{\cal{L}}_j}{2}}\left(X_{j-1}-X_{j+1}\right)$.
  
  \noi The power of the method is that the values of $X_j$ do not
  have to be explicitly calculated, but can be statistically
  estimated. Specifically, the marginalized evidence is obtained
  through the following iterative scheme:
  
  \noi {\bf (1)} the likelihood ${\cal{L}}$ is computed for N
  different points, called active points, which are randomly drawn
  from the prior volume.

  \noi {\bf (2)} the point $X_j$ with the lowest likelihood is found
  and the corresponding prior volume is estimated statistically: after
  $j$ iterations the average volume decreases as $ X_j/X_{j-1}=t $,
  where t is the expectation value of the largest of N numbers
  uniformly distributed between $\left(0,1\right)$.
  
  \noi {\bf (3)} the term
  ${\cal{E}}_j=\frac{{\cal{L}}_j}{2}\left(X_{j-1}-X_{j+1}\right)$ is
  added to the current value of the total evidence;
  
  \noi {\bf (4)} $X_j$ is replaced by a new point randomly
  distributed within the remaining prior volume and satisfying the
  condition ${\cal{L}} >  {\cal{L}}^* \equiv {\cal{L}}_j$;
  
  \noi {\bf (5)} the above steps are repeated until a stopping
  criterion is satisfied.
  
  \noi By climbing up the iso-likelihood surfaces, the method, in
  general, find and quantifies the small region in which the bulk
  of the evidence is located. 

  \noi Different stopping criteria can be chosen.  Following
  \citet{Skilling04}, we stop the iteration when $j \gg \rm{N}H$,
  where H is minus the logarithm of that fraction of prior mass which
  contains the bulk of the posterior mass.  In practical terms this
  means that the procedure should be stopped only when most of the
  evidence has been included. Given the areas ${\cal{E}}_j$, in fact,
  the likelihood initially increases faster than the widths decrease,
  until its maximum is reached; across this maximum, located in the
  region ${\cal{E}}\thickapprox e^{-H}$, the likelihood flatten off
  and the decreasing widths dominate the increasing
  ${\cal{L}}_j$. Since ${\cal{E}}_j\thickapprox e^{-j/\rm{N}}$, it
  takes $\rm{N}H$ iterations to reach the dominating areas.  These
  $\rm{N}H$ iterations are random and are subjected to a standard
  deviation uncertainty $\sqrt{\rm{N}H}$, corresponding to a
  deviation standard on the logarithmic evidence of $\sqrt{\rm{N}
  H}/ \rm{N}$
  
  \begin{equation}
    {\log \cal{E}}=
    \log\left(\sum_j{{\cal{E}}_j}\right)\mathrm{~~~with~~~}
    \sigma_{\log{\cal E}}=\sqrt{\frac{H}{\rm{N}}}\,.
  \end{equation}
   
    \subsubsection{Posterior probability distributions}

  \noi For the lens parameters, the substructure position and the
  logarithm of the source regularization, priors are chosen to be
  uniform on a symmetric interval around the best values which we have
  determined at the second level of the Bayesian inference. The size
  of the interval being at least one order of magnitude larger than
  the errors on the parameters. In practice, we first carry out a fast
  run of the Nested Sampling with few active points $\rm{N}$, this gives us
  an estimate for the non-linear parameter errors. Using the product
  $2\times N_{\rm dim}\times \sigma_\eta$, where $N_{\rm dim}$ is the
  total number of parameters and $\sigma_\eta$ the corresponding
  standard deviation, we can then roughly enclose the bulk of the
  likelihood (note that this can be double-checked and corrected in
  hindsight, if the posterior probability functions are truncated at
  the prior boundaries). Priors on the parameters are taken in such a
  way that this maximum is fully included in the total integral of the
  marginalized evidence. For the main lens parameters and for the
  regularization constant the same priors are used for model with and
  without substructure. For the substructure mass  a flat prior between
  $M_{\rm min}=4.0\times 10^6M_\odot$ and $M_{\rm
  max}=4.0\times 10^9M_\odot$ is adopted, with the two limits given by N-body
  simulations \citep[e.g.][]{Diemand07b, Diemand07a}. In reality,
  the method does not require the parameters to be well known a
  priori, but limiting the exploration to the best fit region
  sensibly reduces the computational effort without significantly
  altering the evidence estimation. From Bayes theorem we have that
  the posterior probability density $p_j$ is given by
  \begin{equation}
    p_j(t)=
    \frac{{\cal{L}}_j}{2}\left(X_{j-1}-X_{j+1}\right)/{\cal{E}}(t)=w_j/{\cal{E}}(t)\,.
  \end{equation}
  The existing set of points $\left(\bmath\eta, \bmath\lambda
  \right)_1$,..., $\left(\bmath\eta, \bmath\lambda \right)_{\rm N}$
  then gives us a set of posterior values that can be then used to
  obtain mean values and standard deviations on the non-linear
  parameters
  \begin{equation}
    \langle\bmath\eta\rangle=\sum_j{w_j\bmath\eta_j}/\sum_j{w_j}\,,
  \end{equation}
  and similarly for $\bmath\lambda$. These samples also provide a
  sampling of the full joint probability density
  function. Marginalising over this function, the full marginalized
  probability density distribution of each parameters can be determined
  (see Section 5.5).
  
  \section{Testing and calibrating the method}\label{sec:test}
  
  In this section we describe the procedure to test the method
  introduced above and to assess its ability to detect dark matter
  substructures in realistic data sets (e.g. from HST). A set of mock
  data, mimicking a typical Einstein ring, is created. We generate
  fourteen different lens models, of which $\rm L_0$ is purely
  smooth, $\rm L_{1 \le i < 13}$ are given by the superposition
  of the same smooth potential with a single NFW dark matter substructure of
  varying mass and position and $\rm L_{13}$
  contains two NFW dark matter substructures with 
  the same mass but with different positions (See Table \ref{tab:lenses}).
  A first approximate reconstruction of the source and of the lens potential
  is performed by recovering the best non-linear lens parameters
  $\bmath\eta$ and the level of source regularization
  $\lambda_s$. These values are then used for the linear grid-based
  optimization, which provides initial values of the substructure
  position and mass. Three extra runs of the non-linear optimization are then
  performed to recover the best set
  $\left(\bmath\eta_b,\lambda_{s,b}\right)$ for the main lens and the
  best mass and position of the substructure (solely modelled with a
  NFW density profile). Finally by means of the Nested-Sampling
  technique described in Section \ref{sec:nested sampling} we
  compute the marginalized evidence, equation (\ref{equ:evidence_integral}), for
  every model twice, once under the hypothesis of a smooth lens and
  once allowing for the presence of one or two extra mass
  substructures. Comparison between these two models allows us to
  assess whether the presence of substructure in the model improves
  the evidence despite the larger number of free parameters.
  
  \subsection{Mock data realisations}
  
  A set of simulated data with realistic noise is generated from a
  model based on the real lens SLACS J1627$-$0055
  \citep{Koopmans06,Bolton06,Treu06}. We assume the lens to be well
  described by a power-law (PL) profile \citep{Barkana98}. Using the
  optimization technique described in Section (\ref{sec:bayes}) we find
  the best set of non-linear parameters
  $\left(\bmath\eta_b,\lambda_{s,b}\right)$. In particular
  $\bmath\eta$ contains the lens strength $b$, and some of the
  lens-geometry parameters: the position angle $\theta$, the
  axis ratio $f$, the centre coordinates $\bmath x_0$ and the density
  profile slope $q$, $\left(\rho \propto r^{-(2q+1)}\right)$. If
  necessary, information about external shear can be included. The
  best parameters are used to create fourteen different lenses and
  their corresponding lensed images. One of the systems is given by a
  smooth PL model while twelve include a dark matter
  substructure with virial mass $\rm M_{vir}=10^7 \rm M_\odot, 10^8
  \rm M_\odot,10^9 \rm M_\odot$ located either on the lowest surface
  brightness point of the ring $P_0$, on a high surface brightness
  point of the ring $P_1$, inside the ring $P_2$ and outside the ring
  $P_3$ (see Table \ref{tab:lenses}). The fourteenth lens
  contains two substructures each with a mass of $\rm M_{vir}=10^8  M_\odot$,
  located respectively in $P_0$ and $P_1$. The substructures are assumed
  to have a NFW profile
  \begin{equation}
    \rho\left(r\right)={\rho_s}{\left(r_s/r\right)\left[1+\left(r/r_s\right)\right]^{-2}}\,,
  \end{equation}
  where the concentration $c=r_{\mathrm {vir}}/r_s$ and the scaling radius $r_s$
  are obtained from the virial mass using the empirical scaling laws
  provided by \citet{Diemand07b, Diemand07a}. The source has an
  elliptical Gaussian surface brightness profile centred in zero
  \begin{equation}
    s\left(\bmath y\right) = s_0 \exp\left[ - (y_1/\delta y_1)^2 - (y_2/\delta y_2)^2 \right]\,.
  \end{equation}
 We assume $s_0=0.25$, $\delta y_1=0.01$ and $\delta y_2=0.04$. 
  
  \begin{table}
    \begin{center}
      \caption {Non-smooth (PL+NFW) lens models. At each of the $P_i$
	positions a NFW perturbation of virial mass $m_{sub}$ is superimposed
	on a smooth PL mass model distribution.}
      \begin{tabular}{cccc} 
	\hline Lens&$\bmath x_{sub}$ $\left( \mathrm{arcsec}
	\right)$&$m_{sub}$ $\left( M_\odot \right)$\\ \hline $\rm
	L_1$&$P_0= (+0.90 ; +1.19)$&$10^7$\\ $\rm L_2$&&$10^8$\\ $\rm
	L_3$&&$10^9$\\ \\ $\rm L_4$&$P_1= (-0.50 ; -1.00)$&$10^7$\\ $\rm
	L_5$&&$10^8$\\ $\rm L_6$&&$10^9$\\ \\ $\rm L_7$&$P_2 = (-0.10 ;
	-0.60)$&$10^7$\\ $\rm L_8$&&$10^8$\\ $\rm L_9$&&$10^9$\\ \\
	$\rm L_{10}$&$P_3 = (-0.90 ; -1.40)$&$10^7$\\ $\rm
	L_{11}$&&$10^8$\\ $\rm L_{12}$&&$10^9$\\ \\
	$\rm L_{13}$&$P_0$ and $P_1 $&$10^8$\\\hline
      \end{tabular} 
      \label{tab:lenses}
    \end{center}
  \end{table}
  
  \subsection{Non-linear reconstruction of the main lens}
  
  We start by choosing an initial parameter set $\bmath\eta_{0}$ for
  the main lens, which is offset from $\bmath\eta_{\rm true}$ that we
  used to create the simulated data. Assuming the lens does not
  contain any substructure we run the non-linear procedure described
  in Section (\ref{sec:bayes}) and optimize $\{\bmath\eta,\lambda_{s}\}$
  for each of the considered systems. At every step of the
  optimization a new set $\{\bmath\eta_i,\lambda_{s,i}\}$ is obtained
  and the corresponding lensing operator $\mathbf{M_c}(\bmath
  \eta_{i},\lambda_{s,i})$ has to be re-computed. The images are
  defined on a 81 by 81 pixels $\left(N_d= 6561\right)$ regular
  Cartesian grid while the sources are reconstructed on a Delaunay
  tessellation grid of $N_s= 441$ vertices. The number of image
  points, used for the source grid construction, is effectively a form
  of a prior and the marginalized evidence (equation \ref{equ:evidence_integral}) can be used to
  test this choice. To check whether the number of image pixels used 
  can affect the result of our modelling, we consider the smooth lens
  $\rm L_0$ and  perform the non-linear reconstruction using one pixel every sixteen, nine, four and
  one. In each of the considered cases we find that the lens parameters are within the relative errors (see Table ~3).
 This suggests that, for this particular case, the choice of number of pixels is not influencing the quality of the reconstruction.  
  In real systems, the dynamic range of the lensed images could be much
  higher and a case by case choice based on the marginalized evidence has to be considered. 
  In Fig. \ref{fig:best1_upr}, the  residuals relative to the system $\rm L_1$ are shown; the noise
  level is in general reached and only small residuals are observed at 
  the position of the substructure. 
  Whether the level of such residuals and therefore the relative detection 
  of the substructure are significant is an issue we will address later on in 
  terms of the  total marginalized evidence.
  
  \subsection{Linear reconstruction: substructure detection}\label{sec:linear rec}
  
\begin{figure}
    \begin{center} 
      \includegraphics[width=8cm]{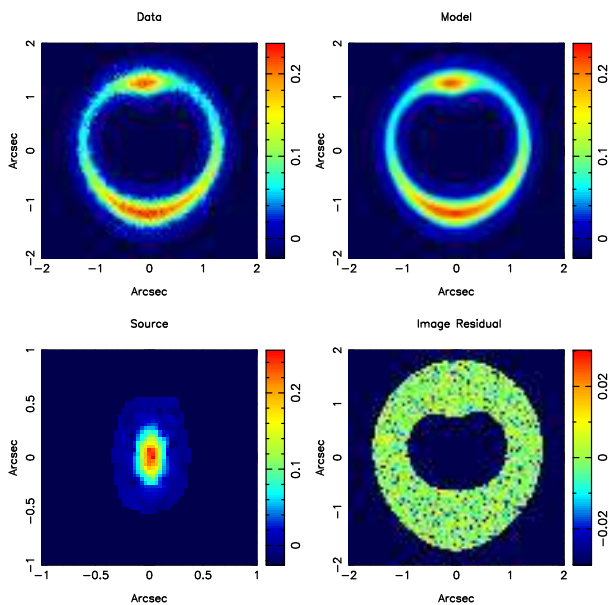}
      \caption{Results of the non-linear optimization for the smooth
	lens $\rm {L_0}$. The top-right panel shows the original mock
	data, while the top-left one shows the final
	reconstruction. On the second row the source reconstruction
	(left) and the image residuals (right) are shown.}
      \label{fig:best_smooth} 
    \end{center}     
  \end{figure}

  The non-linear optimization provides us with a first good
  approximate solution for the source and for the smooth component of
  the lens potential. While this is a good description for the smooth
  model $\rm L_0$ (see Fig. \ref{fig:best_smooth}), the residuals
  (e.g. Fig. \ref{fig:best1_outside_01}) for
  the perturbed model $\rm L_{i\ge1}$ indicate that the
  \emph{no-substructure} hypothesis is improbable and that
  perturbations to the main potential have to be considered. If the
  perturbation is small, this can be done by temporarily assuming that
  $\bmath{\eta}_{i=1}$ reflects the true mass model distribution for the
  main lens and reconstruct the source and the potential correction by
  means of equation (\ref{equ:src_pot_penalty_bayes}). In order to
  keep the potential corrections in the linear regime, where the
  approximation (\ref{equ:src_pot_penalty_bayes}) is valid, both the
  source and the potential need to be initially over-regularised:
  $\lambda_s=10\,\lambda_{s,1}$ and
  $\lambda_{\delta\psi}=3.0\times10^5$ \citep{Koopmans05,
  suyu206}. For each of the possible substructure positions we
  identify the lowest-mass-substructure we are able to recover. In the
  two most favourable cases, $\rm L_1$ and $\rm L_4$, in which the
  substructure sits on the Einstein ring a perturbation of $10^7 \rm
  M_\odot$ is readily reconstructed. For these two positions higher
  mass models, with the exception of $\rm L_2$, will not be further analysed. The systems $\rm
  L_{7,8,9}$ and $\rm L_{10,11,12}$, in which the substructure is
  located, respectively, inside and outside the ring, represent more
  difficult scenarios. In the first case all perturbations below $10^9
  \rm M_\odot$ can be mimicked by an increase of the mass of the main
  lens within the ring, while in the second case these cannot be
  easily distinguished from an external shear effect. For the models
  $\rm L_{1,2,4,9,12}$ convergence is reached after 150 iterations and
  the perturbations are recovered near their known position (Figs. 8 and 9). The grid
  based potential reconstruction indeed leads to a good first
  estimation for the substructure position.

  
  \subsection{Non-linear reconstruction: main lens and substructure}\label{sec:non-linear rec}
  
  In order to compare with numerical simulations, the mass of the
  substructure is required. Performing this evaluation with a grid
  based reconstruction is more complicated and requires some
  assumptions (e.g.\ aperture). To alleviate this problem we assume a
  parametric model, in which the substructures are described by a NFW
  density profile, and we recover the corresponding non-linear
  parameters, mass and position, using the non-linear Bayesian
  optimization previously described.

  \noi To quantify the mass and position of the substructure and to
  update the non-linear parameters when a substructure is added, we
  adopt a multi-step non-linear procedure that relatively fast
  converges to a best PL+NFW mass model. At this level, we neglect the
  smooth lens $\rm L_0$, for which a satisfactory model already has
  been obtained after the first non-linear optimization, and the
  perturbed models $ \rm L_{7,8,10,11}$ for which the substructure
  could not be recovered. We proceed as follows:
  
  \medskip
  
  \noi {\bf (i)} we fix the main lens parameters to the best values
  found in Section (\ref{sec:linear rec}),
  $\{\bmath\eta_1,\lambda_{\rm s,1}\}$. We set the substructure
  mass to some guess value. We optimize for the substructure position
  $\bmath x_{\rm sub,1}$.
  
  \noi {\bf (ii)} we fix $\{\bmath\eta_1,\lambda_{s,1}\}$ and
  the source position $\bmath x_{\rm sub,1}$. We optimize for the
  substructure mass $m_{\rm sub,1}$.
  
  \noi {\bf (iii)} we run the non-linear procedure described in
  Section (\ref{sec:bayes}) by alternatively optimising for the main
  lens, source, and substructure parameters and for the level of source
  regularization. 

  \medskip
  
\begin{figure*}
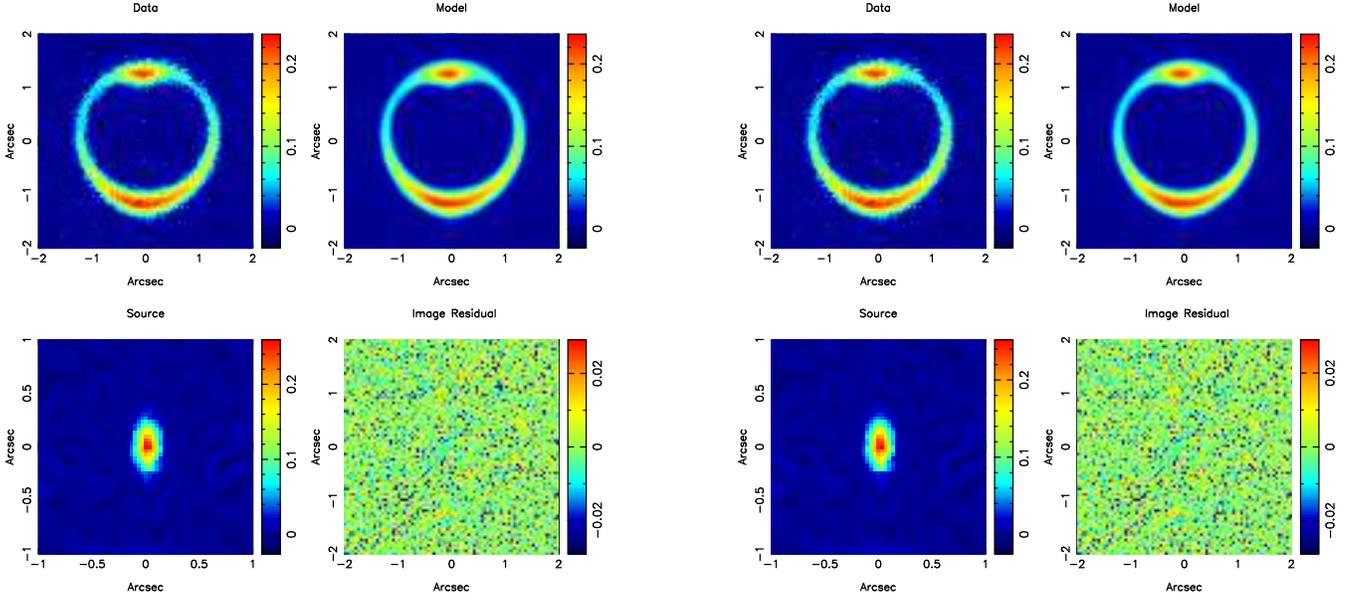

    \begin{center} 
      \includegraphics[width=0.45\hsize]{fig5a}
      \hfill
      \includegraphics[width=0.45\hsize]{fig5b}
      \caption{{\bf Left panel:} Results of the first non-linear
	reconstruction for the smooth component of the perturbed lens
	L$_1$. The top-right panel shows the original mock
	data, while the top-left one shows the final
	reconstruction. On the second row the source reconstruction
	(left) and the image residuals (right) are shown. {\bf Right
	panel:} Final results of the non-linear reconstruction for the
	perturbed lens L$_1$. The top-right panel shows the
	original mock data, while the top-left one shows the final
	model reconstruction obtained after a non-linear optimization
	involving the lens parameters and the substructure position
	and mass. The recovered source is plotted in the low-left
	panel. Image
	residuals (right) are shown.}
      \label{fig:best1_upr} 
    \end{center}
  \end{figure*}

\begin{figure*}
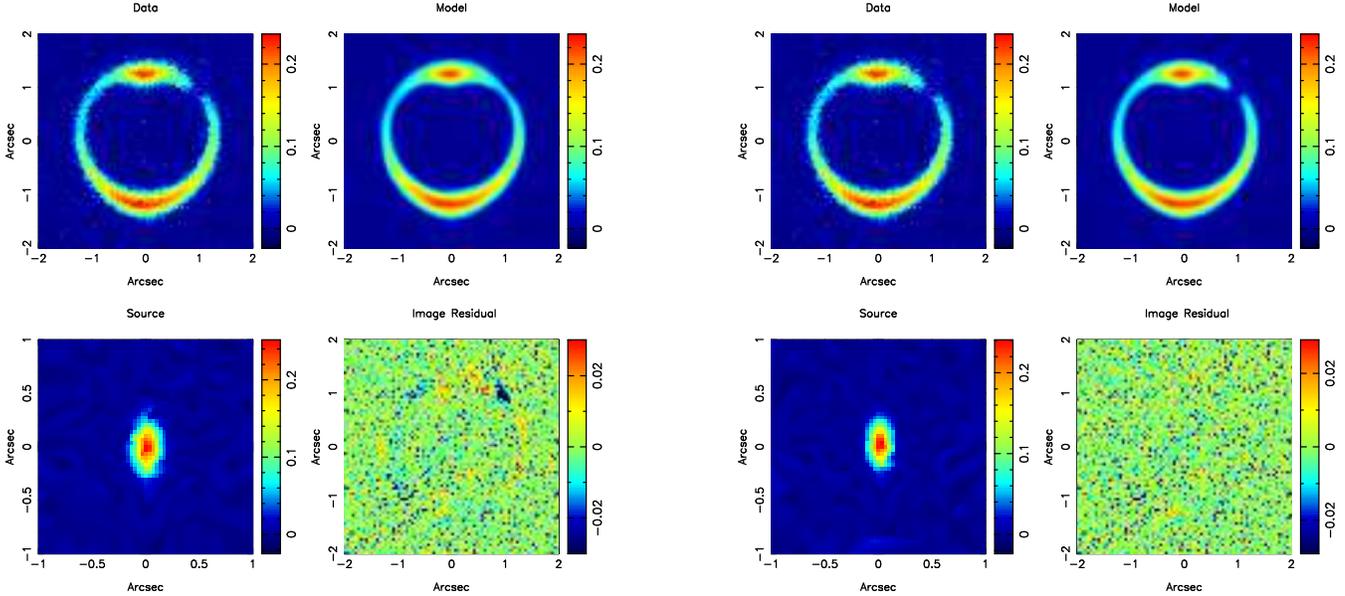

    \begin{center} 
      \includegraphics[width=0.45\hsize]{fig6a}
      \hfill
      \includegraphics[width=0.45\hsize]{fig6b}
      \caption{Similar as Figure~\ref{fig:best1_upr} for L$_2$.}
      \label{fig:best1_upr_001} 
    \end{center}
  \end{figure*}

\begin{figure*}
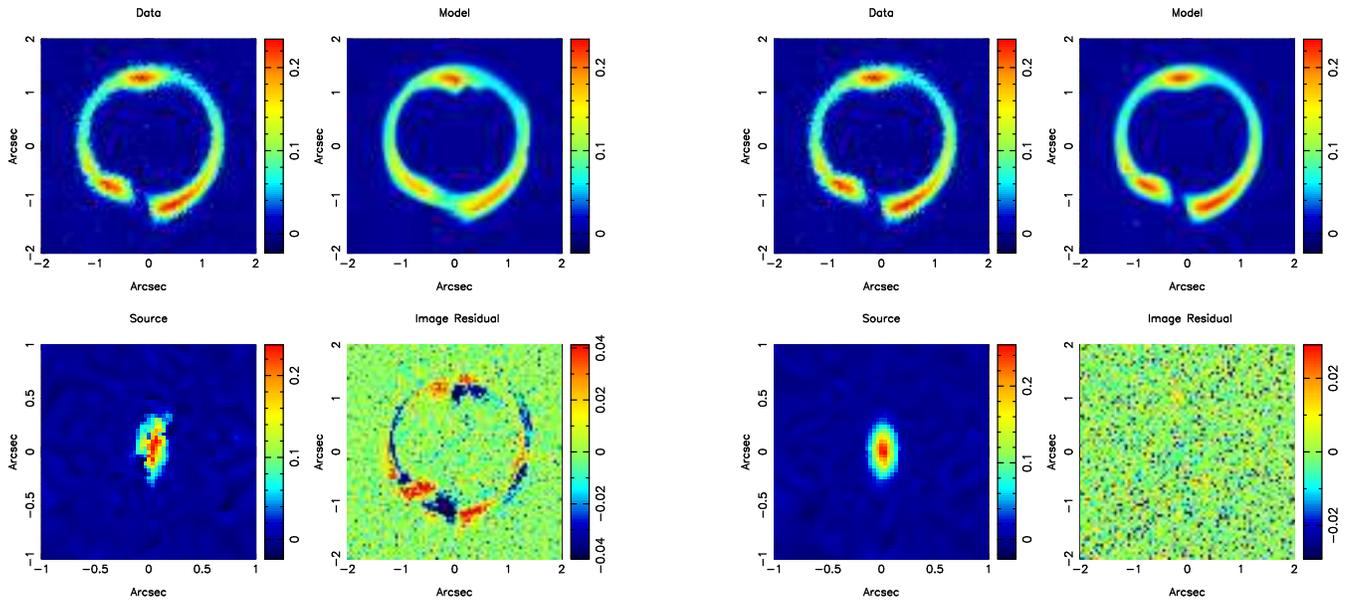

    \begin{center} 
      \includegraphics[width=0.45\hsize]{fig7a}
      \hfill
      \includegraphics[width=0.45\hsize]{fig7b}
      \caption{ Similar as Figure~\ref{fig:best1_upr} for L$_{12}$.}
       \label{fig:best1_outside_01} 
    \end{center}
  \end{figure*}

  \noi This leads to a new set of parameters, $\{\bmath\eta_{\rm b},
  \lambda_{\rm s,b}, m_{\rm sub,b}, \bmath x_{\rm sub,b}\}$. Final
  results for the considered models are listed in
  Table 3 and the
  relative residuals are shown in the Figs. \ref{fig:best1_upr}-\ref{fig:best1_outside_01}, respectively. For all the considered lenses the final
  reconstruction converges to the noise level. 
  

   \subsection{Multiple substructures}
   The lens system $\rm L_{13}$ represents a more complex case in which two substructures
   are included. In particular we are interested in testing
   whether both substructures are detectable and whether their effect may be hidden by the 
   presence of external shear. As for the previously considered cases, we first perform a non-linear 
   reconstruction of the main lens parameters assuming a single PL mass model.
   For this particular system we also include the strength $\rm \Gamma_{sh}$ and the position
   angle $\rm \theta_{sh}$ of the external shear as free parameters. Results for this first step of the reconstruction 
   are shown in Fig. \ref{fig:linear_a}. We then run the linear potential
   reconstruction. One of the two substructures is detected although a significant
   level of image residuals is left (Fig. \ref{fig:best1_sub_double}). 
   The combined effect of external shears ($\rm \Gamma_{sh}=-0.031$) and the substructure in $P_1$
   is not sufficient to explain the perturbation generated by the second substructure at the lowest surface 
   brightness point of the Einstein ring. We therefore include a NFW substructure in 
   the recovered position and run a non-linear reconstruction for the new PL+NFW model, 
   Fig. \ref{fig:linear_b}. We are then able to detect also the second substructure, Fig. \ref{fig:best2_sub_double}. 
   Finally we run a global non-linear reconstruction for the
   PL+2NFW model (Fig. \ref{fig:linear_c}), the noise level is reached and the strength of the external shear  is consistent with zero ($\rm \Gamma_{sh}=0.0001$).
      
  \subsection{Nested sampling: the evidence for substructure}

  When modelling systems as $\rm L_{0}$ or $\rm L_{i\ge1}$ we assume
  that the best recovered values, under the hypothesis of a single
  power-law, provide a good description of the true mass distribution
  and that any eventually observed residual could be an indication for
  the presence of mass substructure. Model comparison within the
  framework of Bayesian statistics gives us the possibility to test
  this assumption. 
  
  \subsubsection{Marginalized Bayesian evidence}
  
  In order to statistically compare two models the
  marginalized evidence (\ref{equ:evidence_integral}) has to be
  computed. As described in Section (\ref{sec:nested sampling}) this
  multi-dimensional and non-linear integral can be evaluated using the
  Nested-Sampling technique by
  \citet{Skilling04}. Specifically the two mass models we wish to
  compare are a single PL, M$_0$, versus a PL+NWF
  substructure, M$_1$. The first one is completely defined by the
  non-linear parameters $\left(\bmath \eta, \lambda_s\right)$, while the
  second needs three extra parameters, namely the substructure mass
  and position. For all these parameters prior probabilities have to
  be defined:
  \begin{equation}
    P\left(\eta_i\right)= \left\{
    \begin{array}{ll}
      \text{constant} & {\rm ~~~for~~~} |\eta_{\rm {b},i}-\eta_i|\leq\delta\eta_i \\ &\\ 
      0 & {\rm ~~~for~~~} |\eta_{\rm {b},i}-\eta_i| > \delta\eta_i
    \end{array} 
    \right.
  \end{equation}
  and
  \begin{equation}
    P\left( x_{\rm {sub},i}\right)= \left\{
    \begin{array}{ll}
      \text{constant} & {\rm ~~~for~~~} | x_{\rm {sub,b},i}- x_{\rm {sub},i}|\leq\delta
      x_{\rm {sub},i}\\ &\\ 0 & {\rm ~~~for~~~} | x_{\rm {sub,b},i}- x_{\rm {sub},i}| > \delta
      x_{\rm {sub},i}
    \end{array}
    \right.
  \end{equation}
  
  \noi where the elements of $\delta\eta_i$ and $\delta x_{\rm sub,i}$
  are empirically assessed such that the bulk of the evidence
  likelihood is included \citep[see][]{Skilling04}. The prior on the
  substructure mass is flat between the lower and upper mass limits
  given by numerical simulations \citep[e.g.][]{Diemand07b,
  Diemand07a}.  Given the lenses $\rm L_{0,1,2,4,9,12,13}$ we run the
  Nested Sampling twice, once for the single PL model and
  once for the PL+NFW (+NFW) one. The two marginalized evidences with
  corresponding numerical errors can be compared from Table ~2. Despite a certain number of authors suggest
  the use of Jeffreys' scale \citep{Jeffreys61} for model comparison, we adopt here a
  more conservative criterion. In particular, we note that the
  perturbed model M$_1$ for the lens system $\rm L_0$ is basically
  consistent with a single smooth PL model M$_0$, with $\Delta{\cal
  {E}}\sim 7.85$. The Bayesian factor for the system $\rm L_4$ is of
  the order of $\Delta{\cal {E}} \sim 21.5$ in favour of the smooth
  model M$_0$, indicating that the detection of such a low-mass
  substructure can formally not be claimed at a significant level. The
  reason why we think this substructure is clearly visible in the
  grid-based results, is that this particular solution is the
  maximum-posterior (MP) solution, whereas the evidence gives the
  integral over the entire parameter space. This implies that there
  must be many solutions near the MP solution that do not show the
  substructure. This indicates that our approach of quantifying the
  evidence for substructure is very conservative.  On the other hand
  the Bayes factor for the lens $\rm L_1$, $\Delta{\cal {E}} = -17.1
  $, clearly shows that the detection of a $10^7 M_\odot$ substructure
  can be significant when the latter is located at a different
  position on the ring. Finally all higher mass perturbations are
  easily detectable independently of their position relative to the
  image ring; Bayes factor for $\rm L_2$, $\rm L_9$, $\rm L_{12}$ and $\rm L_{13}$
  are, in fact, respectively $\Delta{\cal {E}} = -213.0 $,
  $\Delta{\cal {E}} = -2609.7$, $\Delta{\cal {E}} = -4603.4$ and $\Delta{\cal {E}} = -1835.7$.
  Substructure properties for these systems are also confidently
  recovered.
  The main difference between Jeffreys' scale and our criterion for 
  quantifying the significance level of the substructure detection is observed 
  for the system  $\rm L_1$.  If we had to adopt Jeffreys' scale in fact, such detection
  would have to be claimed decisive while we think it is only significant. 
  
\subsection{Posterior probabilities}

  As discussed in Section (\ref{sec:nested sampling}) an interesting
  by-product of the Nested-Sampling procedure is an exploration of the
  posterior probability (\ref{equ:posterior_2}) which provides us with
  statistical errors on the model parameters, see Tables 3 and 4. The
  relative posterior probabilities for $\rm L_0$, $\rm L_1$ and $\rm
  L_2$ are plotted in Fig.~\ref{fig:smooth_weights},
  Fig.~\ref{fig:pert0001_weights} and
  Fig.~\ref{fig:pert001_weights} respectively.  Lets start by
  considering the lens system $\rm L_0$ and the relative probability
  distribution for the substructure mass. Although the model M$_1$, in
  terms of marginalized evidence, is consistent with the single smooth
  PL model M$_0$, there is a small probability for the presence of a
  substructure with mass up to few $10^8 M_\odot$ located as far as
  possible from the ring.  The effect of such objects on the lensed
  image would be very small and could be easily hidden by introducing
  artificial features in the source structure, as suggested by the
  posterior distributions for the source regularization constant.
  This means, that from the image point of view, a smooth single PL
  model and a perturbed PL+NWF with a substructure of $10^8 M_\odot$,
  located far from ring, are not distinguishable from each other as
  long as the effect of the perburber can be hidden in the structure
  of the source. From a probabilistic point of view, however, the second
  scenario is more unlikely to happen.  A similar argument can be
  applied to the lens $\rm L_1$ for which a strong degeneracy between
  the mass and the position of the substructure is observed.  We
  conclude therefore that, although this substructure can be detected
  at a statistically significant level, its mass and position cannot
  be confidently assessed yet.  In contrast, for systems such as $\rm
  L_{2,9,12}$, the effect of the substructure is so strong that it can
  not be mimicked by the source structure or by a different
  combination of the substructure parameters. For these cases not only
  the detection is highly significant, but the properties of the
  perturber can be confidently constrained with minimal biases.
   
  \begin{table}
    \begin{center}
      \caption{marginalized evidence and corresponding standard
	deviation as obtained via the Nested-Sampling
	integration. Results are shown for the hypothesis of a smooth
	lens (PL) and the hypothesis of a clumpy lens potential
	(PL+NFW).}
      \begin{tabular}{cccc} 
      	\hline Lens&Model& $\log {\cal E} \,$&$\sigma_{{\log {\cal E}
	}}\,$\\ \hline $\rm L_0$ & PL & 26332.70&0.33\\ &
	PL+NFW &26324.85&0.30\\ \\ $\rm L_1$ & PL
	&20366.86&0.34\\ &PL+NFW&20383.95&0.30\\ \\ $\rm L_4$
	& PL &20292.40&0.33\\ & PL+NFW &20270.87& 0.29\\ \\
	$\rm L_9$ & PL &17669.41&0.45\\ & PL+NFW
	&20279.13&0.36\\ \\ $\rm L_{12}$ & PL
	&15786.91&0.33\\ & PL+NFW
	&20390.35&0.37\\ \\ $\rm L_{13}$ & PL
	&18509.76&0.24\\ & PL+2 NFW
	&20346.48&0.49\\ \hline
      \end{tabular} 
      \label{tab:evidence}
    \end{center}
  \end{table}
      
 
      
  \section{Conclusions and Future work }
  
  We have introduced a fully Bayesian adaptive method for objectively
  detecting mass substructure in gravitational lens galaxies. The
  implemented method has the following specific features:

  \begin{itemize}

  \item Arbitrary imaging data-set defined on a regular grid can be
    modelled, as long as only lensed structure is included. The code
    is specifically tailored to high-resolution HST data-sets with a
    compact PSF that can be sampled by a small number of pixels.

  \item Different parametric two-dimensional mass-models can be used,
    with a set of free parameter $\bmath \eta$. Currently, we have
    implemented the elliptical power-law density models from
    \citet{Barkana98}, but other models can easily be included.
    Multiple parametric mass models can be simultaneously optimized.
    
  \item A grid-based correction to the parametric potential can
    iteratively be determined for any perturbation that can not easily
    be modelled within the chosen family of potential models (e.g.\
    warps, twists, mass-substructures, etc.).

  \item The source surface-brightness structure is determined on a
    fully adaptive Delaunay tessellation grid, which is updated with
    every change of the lens potential.

  \item Both model-parameter optimization and model ranking are fully
    embedded in a Bayesian framework. The method takes special care not
    to change the number of degrees of freedom during the
    optimization, which is ensured by the adaptive source grid. Methods
    with a fixed source surface-brightness grid can not do this.
    
  \item Both source and potential solutions are regularised, based on
    a smoothness criterion. The choice of regularization can be
    modified and the level of regularization is set by Bayesian
    optimization of the evidence. The data itself determine what
    level of regularization is needed. Hence overly smooth or overly
    irregular structure is automatically penalised.

  \item The maximum-posterior and the full marginalized probability
    distribution function of {\sl all} linear and non-linear
    parameters can be determined, marginalized over all other 
    parameters (including regularization). Hence a full exploration
    of {\sl all} uncertainties of the model is undertaken.

  \item The full marginalized evidence (i.e.\ the probability of the
    model given the data) is calculated, which can be used to rank
    {\sl any} set of model assumptions (e.g. pixel size, PSF) or model
    families. In our case, we intend to compare smooth models with
    models that include mass substructure. The marginalized evidence
    implicitly includes Occam's razor and can be used to assess whether
    substructure or any other assumption is justified, compared to a
    null-hypothesis.
  
  \end{itemize}

  \noi The method has been tested and calibrated on a set of
  artificial but realistic lens systems, based on the 
  lens system SLACS J1627$-$0055. 

  \noi The ensemble of mock data consists of a smooth PL lens and
  thirteen clumpy models including one or two NFW substructures.  Different values
  for the mass and the substructure position have been considered.
  Using the Bayesian optimization strategy developed in this paper we are
  able to recover the smooth PL system and all perturbed models with a
  substructure mass $ \ga 10^7 M_\odot$ when located at the lowest
  surface brightness point on the Einstein ring and with a mass $\geq
  10^9 M_\odot$ when located just inside or outside the ring (i.e.\
  their Einstein rings need to overlap roughly).  For all these models
  we have convincingly recovered the best set of non-linear parameters
  describing the lens potential and objectively set the level of
  regularization. 

  \noi Furthermore, our implementation of the Nested-Sampling
  technique provides statistical errors for {\sl all} model parameters
  and allows us to objectively rank and compare different potential
  models in terms of Bayesian evidence, removing as much as possible
  any subjective choices. Any choice can quantitatively be
  ranked. For each of the lens systems we compare a complete smooth PL
  mass model with a perturbed PL+NFW (+NFW) one.  The method here developed
  allows us to solve simultaneously for the lens potential and the
  lensed source. The latter, in particular, is reconstructed on an
  adaptive grid which is re-computed at every step of the
  optimization, allowing to take into account the correct number
  of degrees of freedom.

  \begin{figure*}
    \begin{center} 
      \includegraphics[width=\hsize]{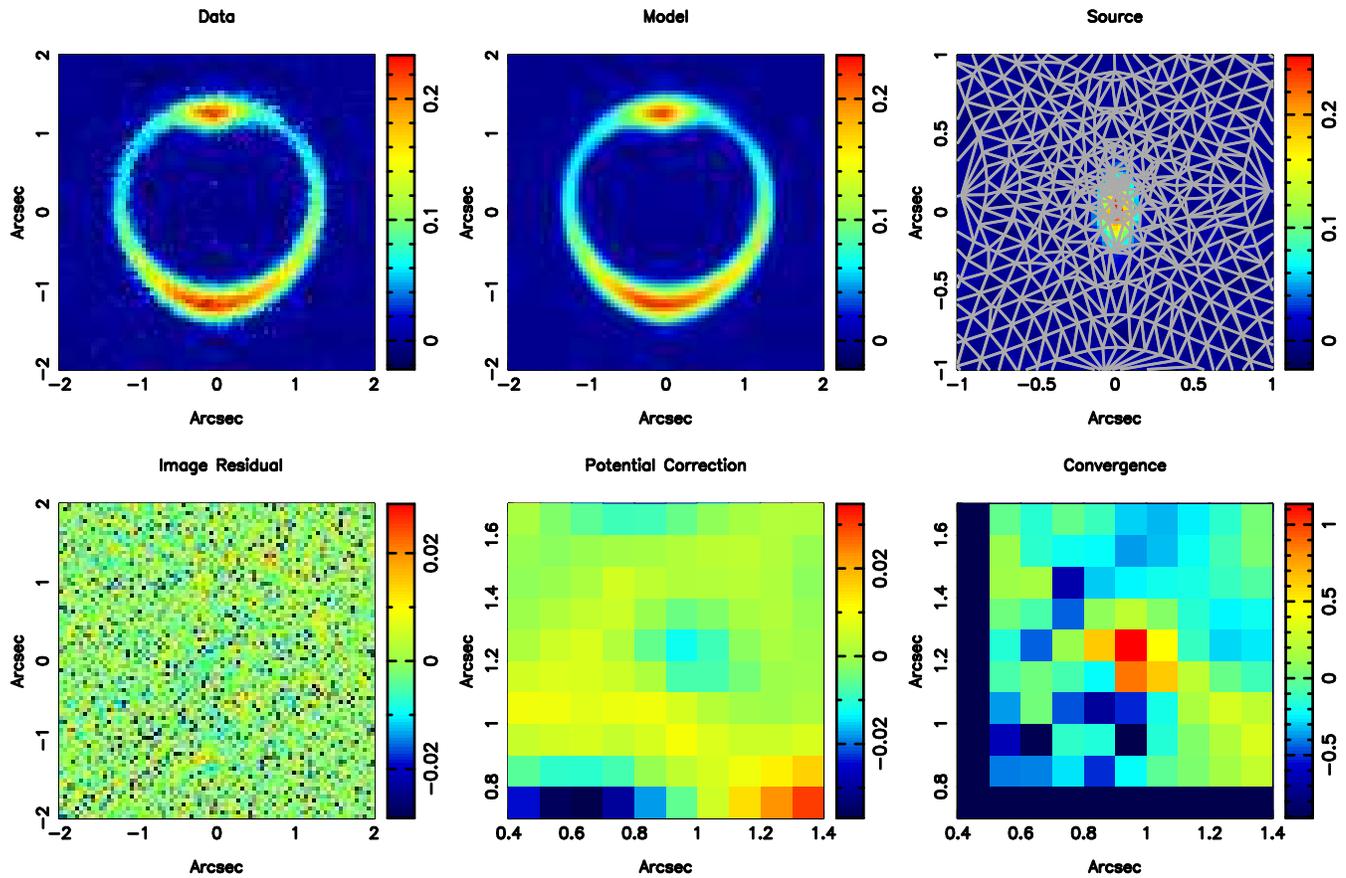}
      \caption{Results of the linear source and potential
	reconstruction for the lens L$_1$. The first row shows
	the original model (left), the reconstructed model (middle)
	and the current-best source, as well as the corresponding adaptive grid. 
	On the second row the image
	residuals (left), the total potential convergence (middle) and
	the substructure convergence (right) are shown. Note 
	that the substructure, although weak, is reconstructed at 
	the correct position.}
      \label{fig:best1_sub_upr} 
    \end{center}
  \end{figure*}
  
   \begin{figure*}
    \begin{center} 
      \includegraphics[width=\hsize]{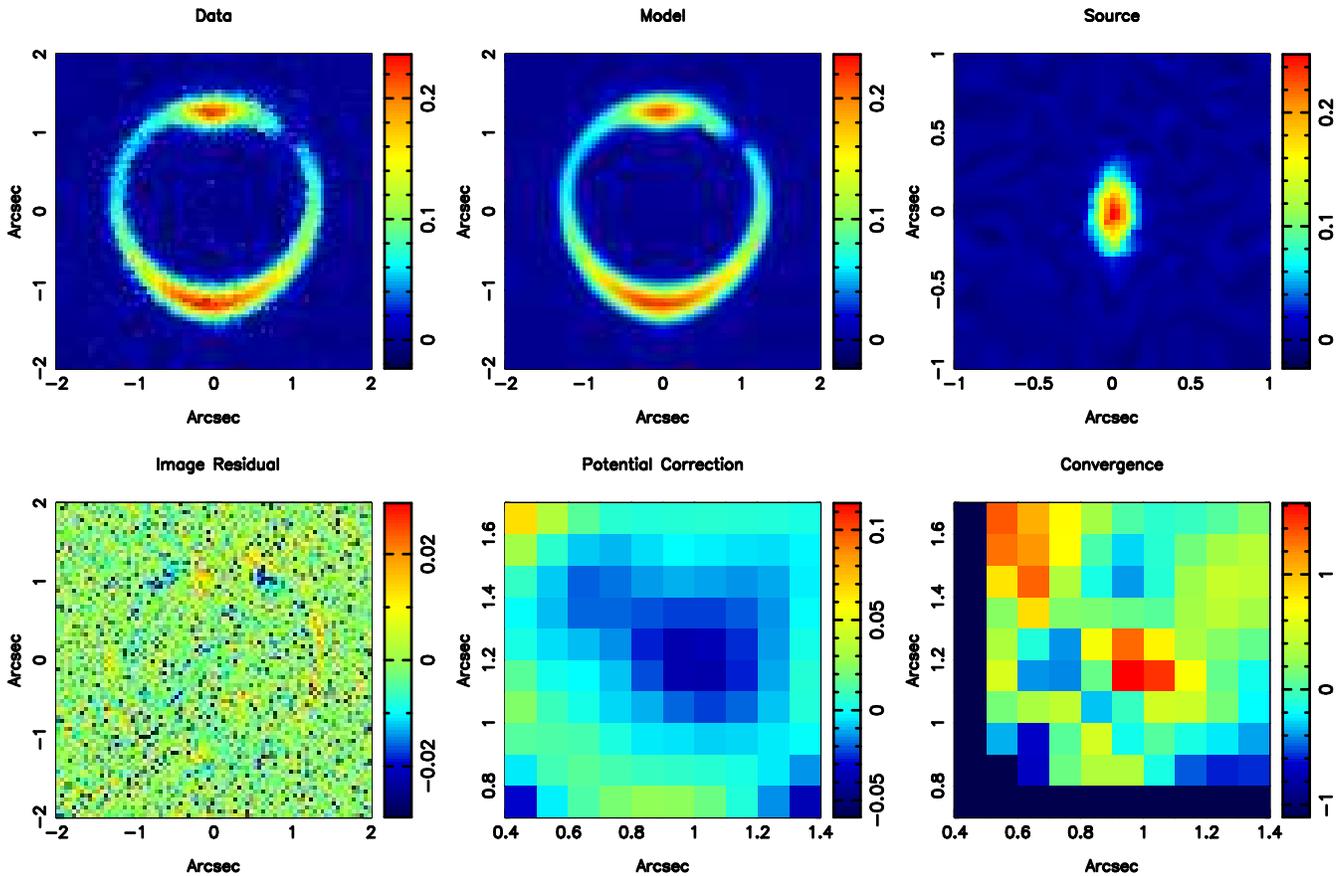}
      \caption{Similar as Figure~\ref{fig:best1_sub_upr} for L$_2$. We note 
	that the substructure is extremely 
	well reconstructed, both at the correct position and in mass.}
      \label{fig:best1_sub_upr_001} 
    \end{center}
  \end{figure*}

   \begin{figure*}
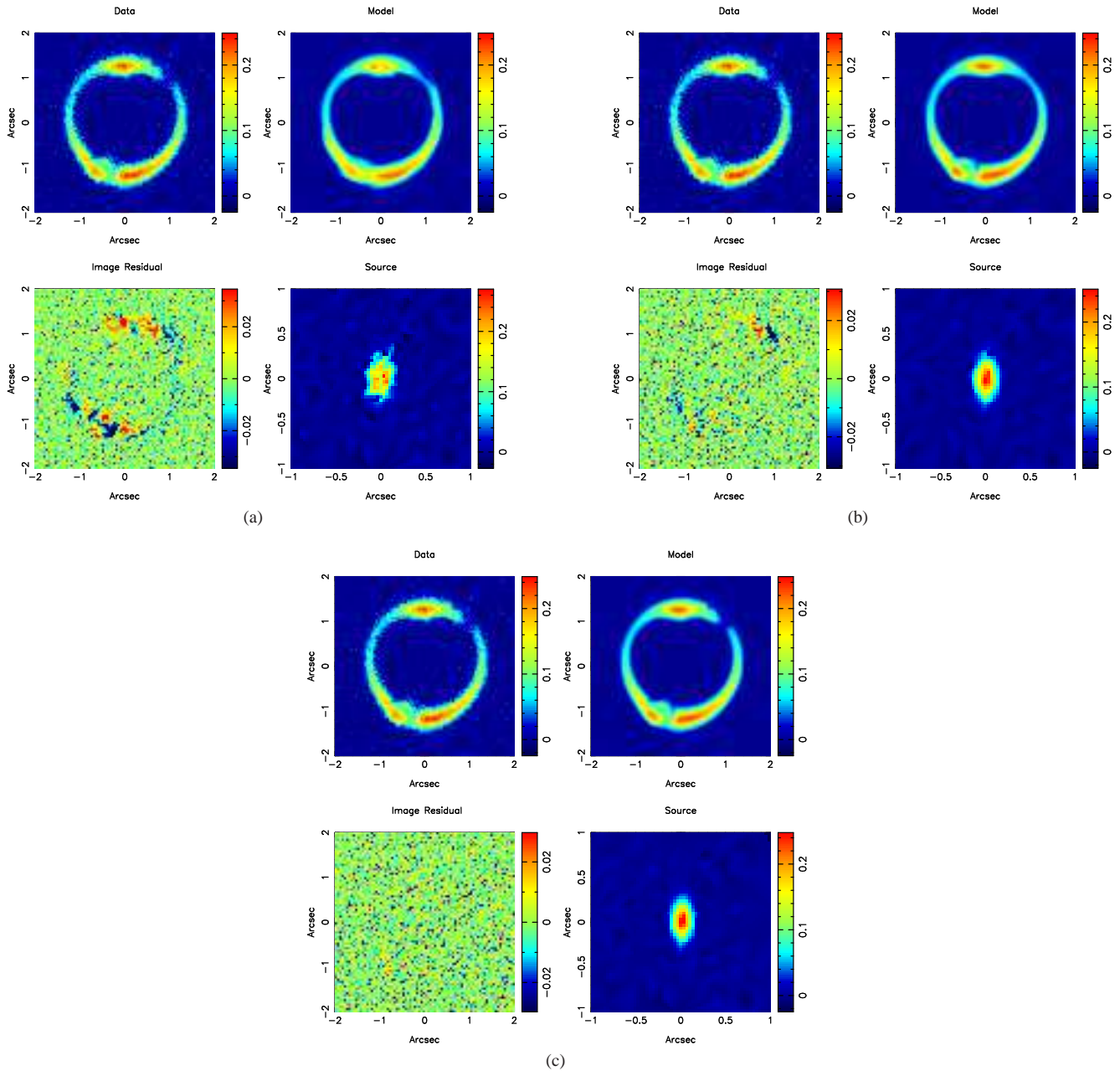

    \begin{center} 
      \subfigure[]{ \includegraphics[width=0.45\hsize]{fig10a}
	\label{fig:linear_a}
      }
      \hfill
      \subfigure[]{ \includegraphics[width=0.45\hsize]{fig10b}
	\label{fig:linear_b}
      }

     \subfigure[]{\centering \includegraphics[width=0.45\hsize]{fig10c}
	\label{fig:linear_c}
      }

      \caption{Non linear reconstruction of the lens $\rm L_{13}$ for a single PL model, a PL+NFW and
      a PL+2NFW one.}
       \label{fig:best_double} 
    \end{center}
  \end{figure*}
  
  \begin{figure*}
    \begin{center} 
      \includegraphics[width=\hsize]{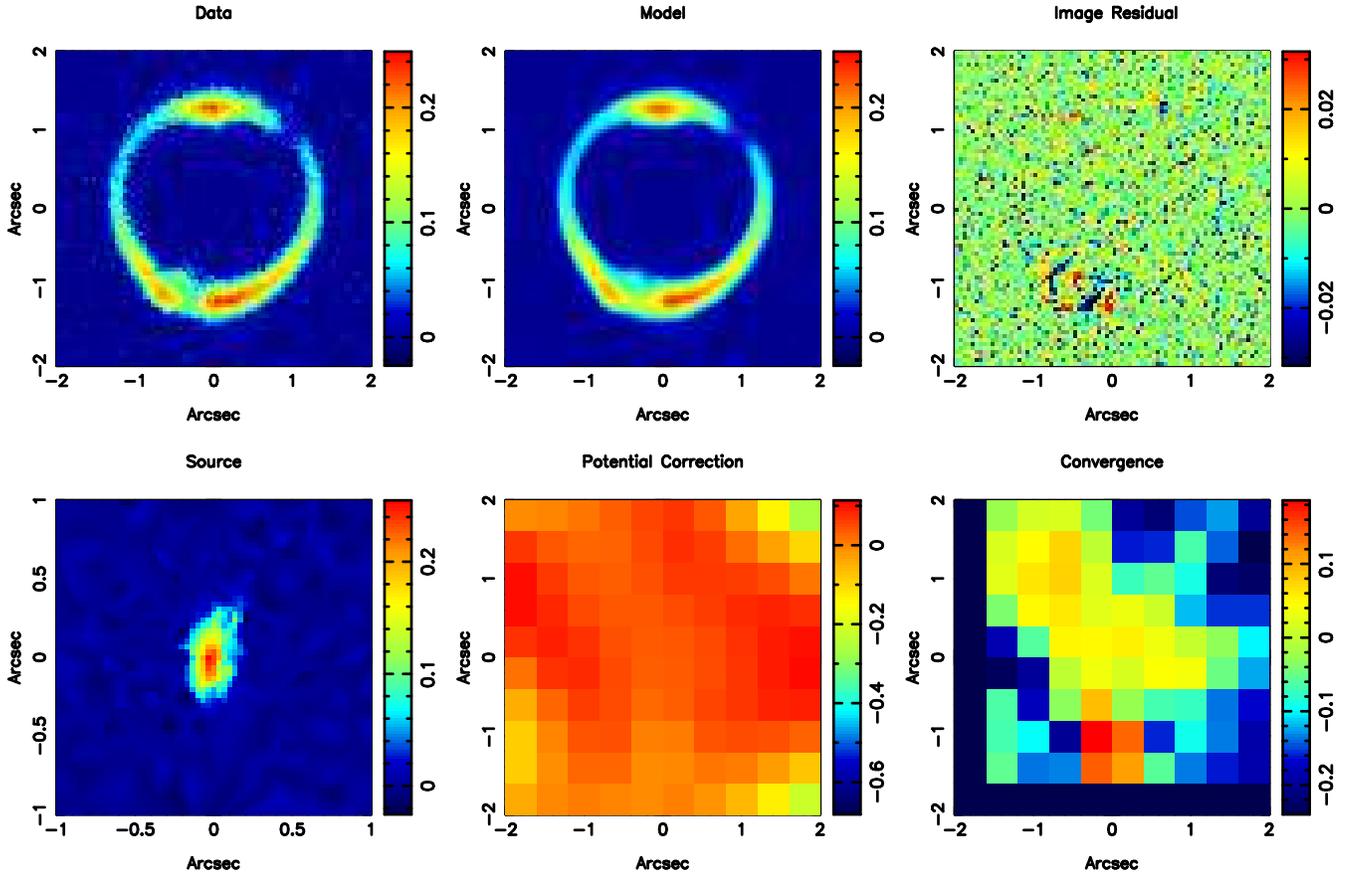}
      \caption{Results of the first linear source and potential
	reconstruction for the lens L$_{13}$. The first row shows
	the original model (left), the reconstructed model (middle)
	and the image residuals. On the second row the current-best source (left), the total potential convergence (middle) and
	the substructure convergence (right) are shown. Note 
	that the substructure, although weak, is reconstructed at 
	the correct position.}
      \label{fig:best1_sub_double} 
    \end{center}
  \end{figure*}
  
 \begin{figure*}
    \begin{center} 
      \includegraphics[width=\hsize]{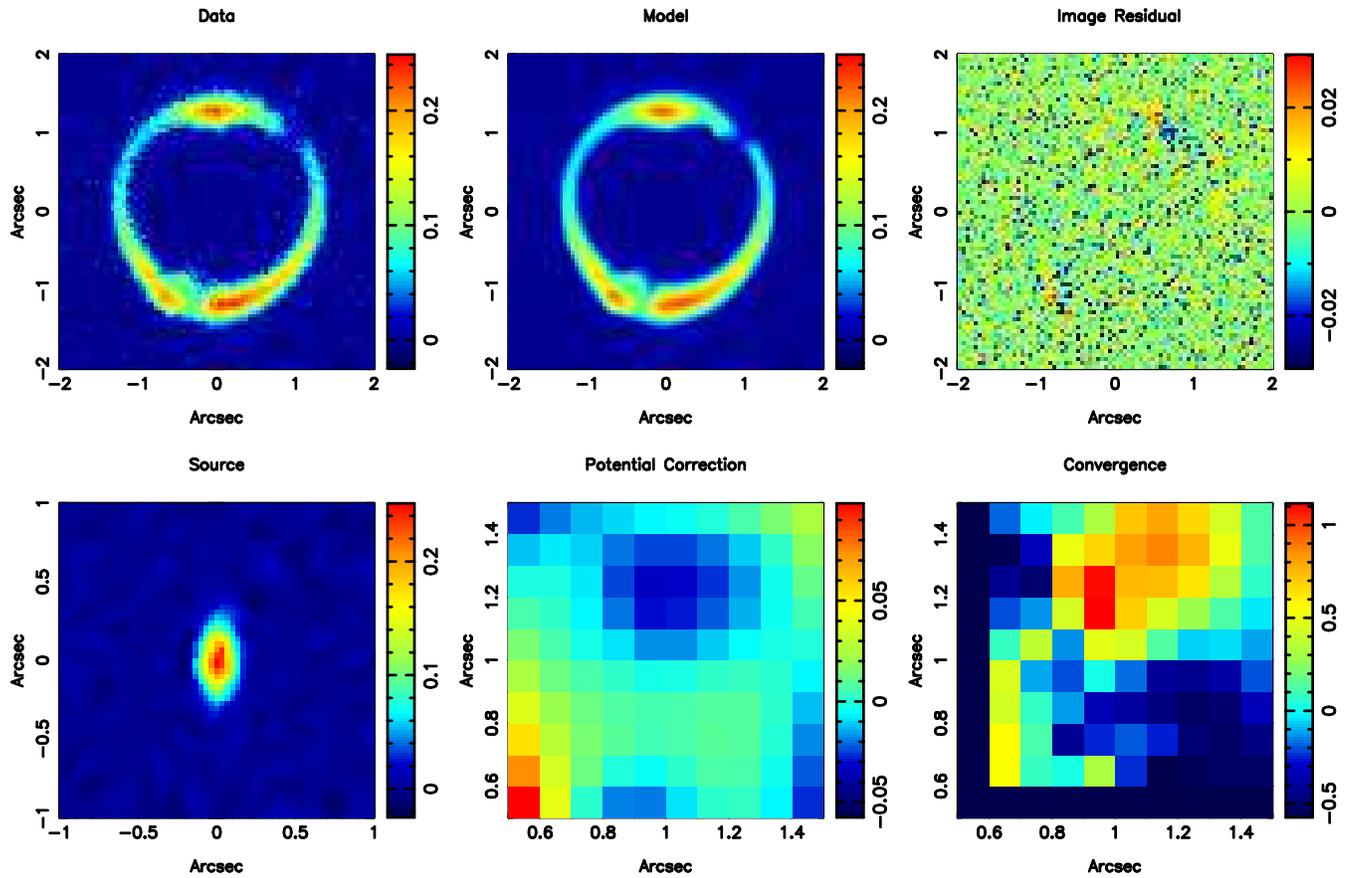}
      \caption{ Results of the second linear source and potential
	reconstruction for the lens L$_{13}$.  }
      \label{fig:best2_sub_double} 
    \end{center}
  \end{figure*}

  \noi In this paper we have considered systems which contains at most two CDM substructures. Although it may appear as a very small 
  number when compared with predictions from N-body simulations within the virial radius, this represents a realistic scenario. 
  As we have shown, our method, with current HST data, is mostly sensitive 
  to perturbations with mass $\ga 10^7\rm M_\odot$ and located on the Einstein ring ($\Delta\theta\sim\mu\theta_{\rm ER}$). 
  The projected volume that we are able to probe is therefore small compared to the projected volume within the virial radius. 
  The probability that more than two substructures have this right combination of mass and position is relatively low and we expect most of the 
  real systems to be dominated by one or at most two perturbers.
  \noi Despite these new results, further improvements can still be
  made. We think, for example, that an adaptive source grid based on surface
  brightness, rather than magnification, or a combination, could be
  more suitable for the scientific problem considered here. 
  
  \noi The method will soon be applied to real systems, as for example
  from the \emph{Sloan Lens ACS Survey} sample of massive early-type galaxies
  \citep{Koopmans06,Bolton06,Treu06}. This will lead to powerful new
  constraints or limits on the fraction and mass distribution of
  substructure. Results will be compared with CDM simulations.

  \section*{Acknowledgements} The authors would like to thank Matteo
  Barnab\`e, Oliver Czoske, Antonaldo Diaferio, Phil Marshall, Sherry Suyu and the anonymous referee  for useful 
  discussions. They also thank the Kavli Institute for Theoretical Physics 
  for hosting the gravitational lensing workshop in fall 2006, during which 
 important parts of  this work were made. SV and LVEK are supported (in part) through an
  NWO-VIDI program subsidy (project number 639.042.505).

  \bibliography{ms}


  \begin{figure*} 
    \begin{center} 
      \includegraphics[width=\hsize]{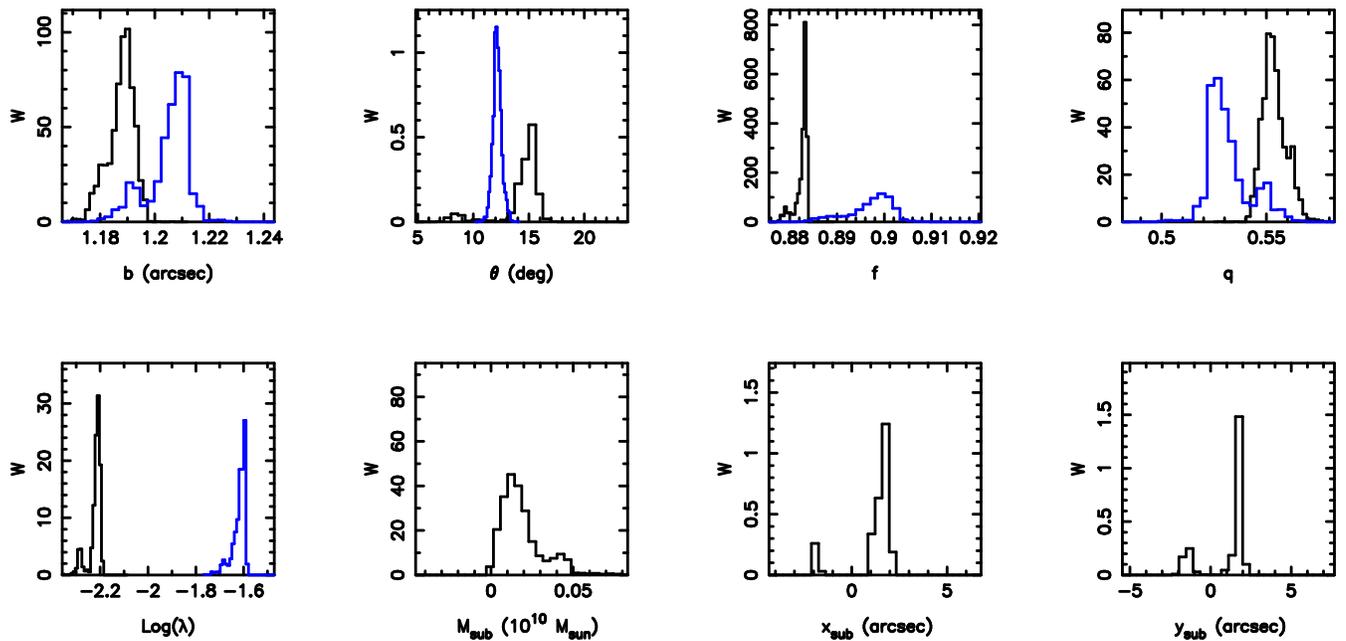}
      \caption{Posterior probability distributions for the non linear
	parameters of the smooth lens model $\rm L_0$ as obtained from
	the Nested-Sampling evidence exploration. In particular
	results for two different models are shown, a smooth PL
	potential (blue histograms) and a perturbed PL+NFW lens
	(black histograms). From up left, the lens strength, the
	position angle, the axis ratio, the slope, the logarithm of
	the source regularization constant, the substructure mass and
	position are plotted.}
      \label{fig:smooth_weights} 
    \end{center}
  \end{figure*}
  
  \begin{figure*} 
    \begin{center} 
      \includegraphics[width=\hsize]{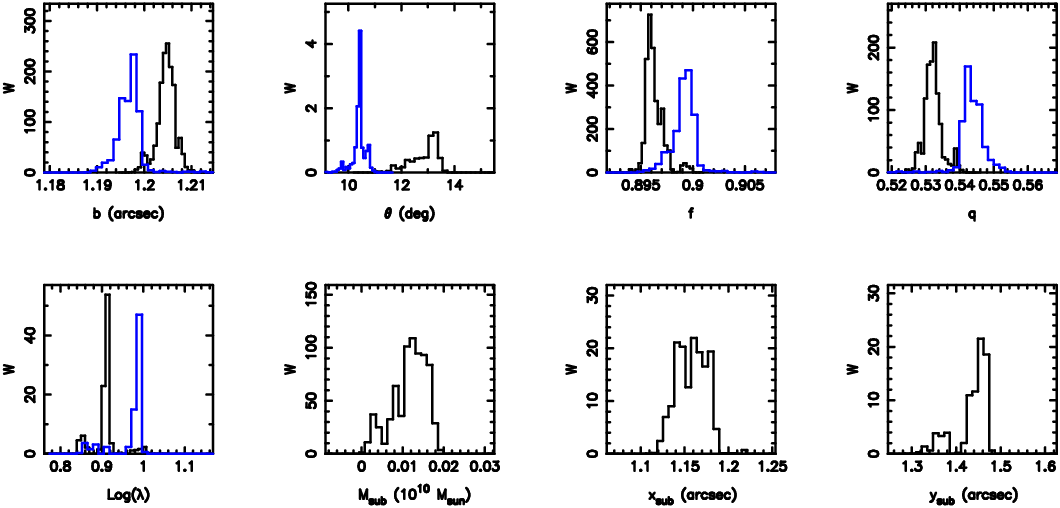}
      \caption{Similar as Figure~\ref{fig:smooth_weights} for L$_1$.}
        \label{fig:pert0001_weights} \end{center}
	\end{figure*} 
	
	\begin{figure*} 
    \begin{center} 
      \includegraphics[width=\hsize]{fig15}
      \caption {Similar as Figure~\ref{fig:smooth_weights} for L$_2$.}
      \label{fig:pert001_weights} 
    \end{center}
  \end{figure*}

 \begin{figure*} 
    \begin{center} 
      \includegraphics[width=\hsize]{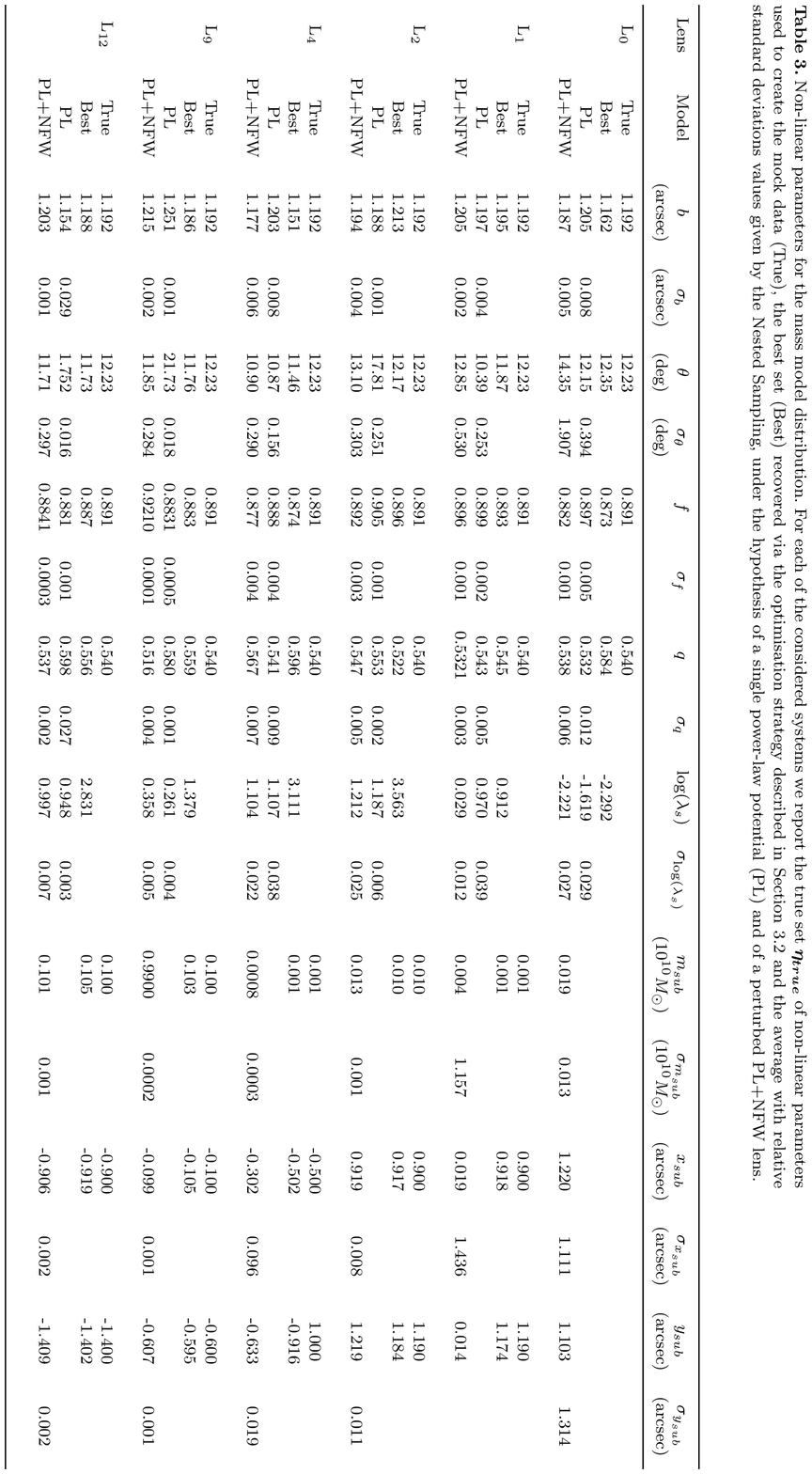}
         \end{center}
  \end{figure*}
  
  \begin{figure*} 
    \begin{center} 
      \includegraphics[width=\hsize]{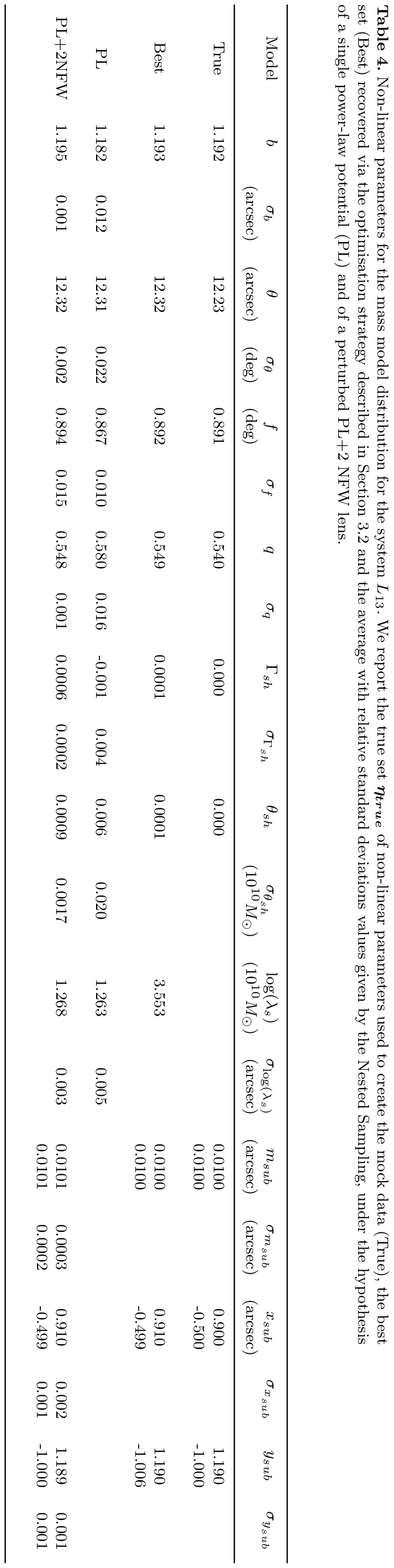}
    \end{center}
  \end{figure*}

\clearpage

\newpage \label{lastpage}

\end{document}